\documentclass[reprint,prd,nofootinbib,superscriptaddress]{revtex4}
\usepackage{CJK}
\usepackage{amsfonts}
\usepackage{amsmath,slashed}
\usepackage{amssymb}
\usepackage[english]{babel}
\usepackage{graphicx}
\usepackage{epsfig}
\usepackage{bm}
\usepackage{verbatim}
\usepackage[utf8]{inputenc}
\usepackage{booktabs}
\usepackage{multirow}
\usepackage{subfig}
\usepackage{slashed}
\usepackage{xcolor}
\usepackage[colorlinks=true,urlcolor=red,citecolor=red]{hyperref}
\usepackage[font=small]{caption}
\usepackage{float}
\usepackage{blindtext}
\usepackage{placeins}

\newcommand{\hs}{\hspace*{0.3cm}}

\newcommand{\be}{\begin{equation}}
	\newcommand{\ee}{\end{equation}}
\newcommand{\bea}{\begin{eqnarray}}
	\newcommand{\eea}{\end{eqnarray}}
\newcommand{\ben}{\begin{enumerate}}
	\newcommand{\een}{\end{enumerate}}
\newcommand{\bit}{\begin{itemize}}
	\newcommand{\eit}{\end{itemize}}
\newcommand{\bde}{\begin{widetext}}
	\newcommand{\ede}{\end{widetext}}

\newcommand{\crn}{\nonumber \\}

\newcommand{\al}{\alpha}
\newcommand{\la}{\lambda}

\newcommand{\ga}{\gamma}

\newcommand{\om}{\omega}
\newcommand{\pa}{\partial}

\newcommand{\fr}{\frac}

\newcommand{\bc}{\begin{center}}
	\newcommand{\ec}{\end{center}}

\newcommand{\ep}{\epsilon}

\newcommand{\eq}{\eqref}
\newcommand{\mathsym}[1]{{}}
\newcommand{\gev}{~\mathrm{GeV}}

\definecolor{bostonuniversityred}{rgb}{0.8, 0.0, 0.0}

\def\gsim{\raise0.3ex\hbox{$\;>$\kern-0.75em\raise-1.1ex\hbox{$\sim\;$}}}
\def\lsim{\raise0.3ex\hbox{$\;<$\kern-0.75em\raise-1.1ex\hbox{$\sim\;$}}}

\newcommand{\hpu}{ Department of Physics, Hanoi Pedagogical University 2, Xuan Hoa, Phu Tho, Vietnam}
\newcommand{\hvkh}{Graduate Univesity of Sience and Technology, Vietnam Academy of Sience and Technology, Hanoi, Vietnam}
\newcommand{\iop}{Institute of Physics, VAST, 10 Dao Tan, Giang Vo, Hanoi, Vietnam}
\newcommand{\ltp}{Bogoliubov Laboratory of Theoretical Physics, Joint Institute for Nuclear Research, Dubna 141980, Russian Federation}
\newcommand{\stai}{ Subatomic Physics Research Group,
	Science and Technology Advanced Institute,\\
	Van Lang University, Ho Chi Minh City, Vietnam}
\newcommand{\steh}{  Faculty of Applied Technology, School of  Technology,  Van Lang University, Ho Chi Minh City, Vietnam}

\begin{document}
	\title{ Heavy neutral bosons and dark matter in the 3-3-1 model with axionlike particle }
	\author{T.T. Hieu $^{a,b}$}
	\email{trantrunghieu@hpu2.edu.vn}
	\author{V.H. Binh$^{c,f}$}
	\email{vhbinh@iop.vast.vn}
	\author{H. N. Long$^{d,e}$}
	\email{hoangngoclong@vlu.edu.vn}	
	\author{H. T. Hung$^{a,f}$}
	\email{hathanhhung@hpu2.edu.vn(corresponding author)}
	\affiliation{
		$^a$ \hpu \\
		$^b$ \hvkh \\
		$^c$ \iop \\
		$^d$ \stai \\
		$^e$ \steh \\
		$^f$ \ltp \\	
	}
	\date{\today }

	\begin{abstract}
		We consider heavy neutral  bosons in the 3-3-1 model with axionlike particles (331ALP), including the Higgs boson and the $Z^\prime$ boson which are outside the standard model (SM). Based on gluon-gluon fusion at the LHC, we investigate the signals of cross-sections in the parameter space region satisfying the current experimental limits of lepton flavor violating decay, including  processes involving both charged leptons and Higgs boson, and provide predictions of $m_{h_2}\geq 630 ~\mathrm{GeV}$. A new gauge boson, labeled as $Z^{\prime}$, has its mass excluded in the smaller  domain $5.1 ~\mathrm{TeV}$ based on investigating the cross-section of the processes that $Z^{\prime}$ mediates combined with experimental limits of the search for high-mass dilepton resonances at ATLAS and CMS. We consider the stability of odd-$Z_2$ particles, with $Z_2$ is assumed a residual symmetry after spontaneous symmetry breaking stages, to point out dark matter candidates in the model. Investigating the relic density of dark matter within experimentally permissible limits, we established a relationship between the mass of dark matter and the breaking scale of axion.
		
	\end{abstract}
	
	\keywords{lepton-flavor-violating decays, CP-even Higgs bosons, gluon-gluon
		fusion at LHC, dark matter, heavy neutral bosons}
	
	\maketitle
	\noindent
	
	\section{\label{intro}Introduction}
	
	 The hypothesis of lepton flavor violating (LFV) processes is increasingly being considered and used widely through the explanations of new physics involved such as: the polarization of $\tau$ and the mixing of charged leptons in two last generations in $\tau \rightarrow 3\mu$ decay at Belle-II \cite{Calcuttawala:2018wgo}, polarization probes of $\tau$ and Hyperon in the quasi-elastic scattering processes \cite{Yan:2025luk}, the influence of LFV interactions mediated by heavy scalars with lepton number 2 (bilepton scalars) on the supernova dynamics \cite{Lychkovskiy:2010ue}, etc. The neutral lepton part is trusted with confirmation of the mass and oscillations of neutrinos \cite{ Patrignani:2016xqp,Tanabashi:2018oca}, and the charged lepton part is also supported with given limits on branching ratios as \cite{ParticleDataGroup:2024cfk,ATLAS:2019erb,Zyla:2020zbs, BaBar:2009hkt, MEG:2016leq}:
	\bea
	\mathrm{Br}(\mu \rightarrow e\gamma)\leq 4.2\times 10^{-13},\,
	\mathrm{Br}(\tau \rightarrow e\gamma)\leq 3.3\times 10^{-8},\,
	\mathrm{Br}(\tau \rightarrow \mu\gamma)\leq 4.4\times 10^{-8},\label{lalb-limmit}
	\eea
	and the lepton flavor violating decays of SM-like Higgs boson (LFVHD) also have experimental limits given as \cite{CMS:2018ipm,ATLAS:2019pmk,ATLAS:2019old}. 
	\bea
	\mathrm{Br}(h \rightarrow \mu\tau)\leq 10^{-3},\,
	\mathrm{Br}(h \rightarrow \tau e)\leq 10^{-3},\,
	\mathrm{Br}(h \rightarrow \mu e)\leq  6.1 \times 10^{-5}. \label{hmt-limmit}
	\eea
	
	The LFV processes are of greater interest in the models  beyond the standard model (BSM), where are abundant sources of lepton flavor violation. They may appear alongside mass generation mechanisms for neutrinos, including two main types: effective operators \cite{Mizukoshi:2010ky, Dias:2005yh, Hung:2022vqx, Hung:2022kmv,PhysRevD.47.2918} and seesaw mechanisms \cite{Boucenna:2015zwa, Hernandez:2013hea, Nguyen:2018rlb, Hung:2021fzb, Ferreira:2019qpf, CarcamoHernandez:2019pmy, Catano:2012kw, Hernandez:2014lpa, Dias:2012xp}. Both these types yielded good results regarding LFVHD, even showing that $Br(h\rightarrow \mu\tau)$ can reach $\mathcal{O}(10 ^ {-4})$ \cite{Hue:2015fbb, Thuc:2016qva, Zhang:2015csm, Herrero-Garcia:2017xdu}. Theoretical results have shown that the signals of LFVHD are much smaller than the experimental limit at the LHC \cite{Blankenburg:2012ex}, even with the contribution of Majorana neutrinos \cite{ Pilaftsis:1992st}. One publication also pointed out that the LFVHD signal at one loop order in BSM is always less than $10^{-4}$, because it is suppressed by a loop factor and constraints from the lepton flavor violating of charged lepton (cLFV) \cite{Herrero-Garcia:2016uab}. Furthermore, some publications have indicated a parameter space region that satisfies the experimental limits of both cLFV and LFVHD, which is well-suited for investigating other phenomena, for example, predicting the signal $(g-2)_\mu$ in the parameter space region satisfying the experimental limits of LFVHD \cite{Hong:2024yhk}, finding the signal of $Br(Z \rightarrow e_ae_b)$ and comparing it with $Br(h \rightarrow e_ae_b)$ \cite{Hong:2024swk}. Morever, to explain oscillation of neutral mesons and the rare decay of mesons, one often creates tree-level quark FCNCs with the appearance of a new neutral gauge boson $Z^\prime$ and a singlet scalar. To accomplish this, $U(1)_N$ symmetry is added the gauge group of SM. An interesting consequence of this model is that it naturally explains the supplement to the measurement of the W boson mass of CDF or CMS.\cite{Duy:2024txy}.\\
	
	Recently, the experimental LHC based on gluon-gluon fusion at a center of mass energy $\sqrt{s}=13~\mathrm{TeV}$ have yielded very noteworthy results. The search for a heavy neutral Higgs boson in the range of $200 \to 900~ \mathrm{GeV}$,  based on the $\mu \tau$ decay mode,  performed by CMS, using a CMS detector with an intergrated luminosity of $35.9 ~\mathrm{fb}^{-1}$, has shown upper limits on production cross section multiplied by the branching ratio to be in the range of $51.9(57.4)~\mathrm{fb}$ to $1.6(2.1)~\mathrm{fb}$ at $95\%~\mathrm{CL} $ \cite{CMS:2019pex}. Furthermore, the search for high-mass resonances in the range of  $250 \to 6000~ \mathrm{GeV}$ has also been carried out by ATLAS ($139 ~\mathrm{fb}^{-1}$) and CMS($140 ~\mathrm{fb}^{-1}$) and an important consequence derived from that is the relationship between the upper limit of cross-section times branching ratio at $95\%~\mathrm{CL} $ with the resonant mass of the heavy neutral boson \cite{ATLAS:2019erb, CMS:2021ctt}. Based on the relationship, one can obtain the mass limits of heavy neutral bosons. This is a promising theoretical suggestion for predicting the masses of heavy neutral bosons through investigating the product of cross-section and the branching ratio in a particular model.

	The 3-3-1 models can offer axions that are not only naturally light but also avoid the quantum gravity effects when introducing the local $Z_N$ ($N=11,~13$) symmetry \cite{Dias:2003iq,Dias:2003zt,Dias:2002gg}. The act of $Z_N$, as part of the gauge invariance, along with particle content, renormalizability, and Lorentz invariance, results in the Peccei-Quinn (PQ) symmetry in the model being automatically implemented in classical Lagrangian. The light mass of axion is resolved by adding a gauge singlet whose VEV is at the axion symmetry breaking scale about $10^{10} ~ \mathrm{GeV}$ in a 3-3-1 model with the gauge group as $\mbox{SU(3)}_C\otimes \mbox{SU(3)}_L\otimes \mbox{U(1)}_X \otimes Z_{11}\otimes Z_2$ \cite{Ferreira:2016uao}. A model established based on the gauge group and particle content as in Ref. \cite{Ferreira:2016uao}, differing only in the act of $Z_2$ on the fields, has given a fairly complete of Higgs sector phenomenology \cite{alp331}, it is called 331ALP in short. The 331ALP possesses many interesting properties as: the physical states of axion like particle and other scalar particles were given, the mass of the scalar sector was indicated in the range of $100 ~ \mathrm{GeV} - 1~ \mathrm{TeV}$ based on numerical analysis, rare top quark decays, meson oscillations, as well as the lepton flavor conserving decay of SM-like Higgs boson were also analyzed. However, LFV processes, the masses of heavy neutral bosons, and dark matter have yet to be considered in 331ALP. In this work, we will address those issues.
	
	The paper is organized as follows. In the next section, we review the model and give masses spectrum of gauge and Higgs bosons. We calculate the Feynman rules and analytic formulas for cLFV and LFVHDs  in Section \ref{hLFV}. $Z^\prime$ boson and dark mater are discussed in Section \ref{heavy} and Section \ref{darkmatter}. Conclusions are in Section \ref{Conclusion}. Finally, we provide Appendix \ref{appen_PV},\ref{appen_loop} to calculate the amplitude and show numerical results of $h_{1,2} \rightarrow \mu \tau$ decays.

	\section{Review of the model}
	\label{model}
	
	\subsection{Particle content and discrete symmetries}
	\label{dis}
	
	The fermion sector of the model and their $\mbox{SU(3)}_C\times \mbox{SU(3)}_L\times \mbox{U(1)}_X$ assignments are:
	\begin{align}
		\psi_{aL} =\left( \nu_{a},l_{a},  (\nu_{aR})^c\right) _{L}^{T}\sim
		\left( 1,3,-1/3\right),\hs l_{aR}\sim \left( 1,1,-1\right),~N_{aR} \sim (1,1,0),\crn
		Q_{3L} =\left( u_3,d_3,U_3\right) _{L}^{T}\sim \left(
		3,3,1/3\right),~ Q_{n  L} =\left( d_{n  },-u_{n	},D_n\right)_{L}^{T}\sim \left( 3,3^{\ast },0\right), \crn
		u_{aR} \sim  \left(3,1,2/3\right),~ U_{3R} \sim \left(3,1,2/3\right),\hs  
		d_{aR}\sim (3,1,-1/3),~ D_{nR} \sim (3,1,-1/3), \label{spectrumofparticles}
	\end{align} 
	where $n  = 1,2$ and $a=\{n,3\}$ are family indices.
	The $U_{3L,R}$ and $D_{nL,R}$ are exotic quarks with charges of $2/3$ and $-1/3$, respectively, the same as the usual u-type and d-type, whereas $N_{aR}$ and $\nu_{aR}$ are right-handed Majorana and Dirac neutrinos, respectively. 
	
	We need three $SU(3)_L$ triplet scalars, $\eta, \rho, \chi$ to generate mass for gauge bosons and fermions as usual in 3-3-1 models with right-handed neutrinos \cite{Dias:2005yh,cl}. However, in this case, we require an additional neutral $SU(3)_L$ singlet scalar, $\phi$, to generate mass for exotic neutrinos. Therefore, the multiplet scalars and their vacuum expectation values (VEVs) are given as follows \cite{alp331,Long:2024rdq,Ferreira:2016uao} :
	\bea
	\eta = \left( \begin{array}{c}
		\eta_1^0  \\
		\eta_2^- \\
		\eta_3^0
	\end{array}
	\right) \sim \left( 1,3,-\frac{1}{3}\right),\,
	\chi = \left( \begin{array}{c}
		\chi_1^0  \\
		\chi_2^- \\
		\chi_3^0
	\end{array}
	\right) \sim \left( 1,3,-\frac{1}{3}\right),\,
	\rho = \left( \begin{array}{c}
		\rho_1^+  \\
		\rho_2^0 \\
		\rho_3^+
	\end{array}
	\right) \sim \left( 1,3,\frac{2}{3}\right),\,\phi  \sim  (1,1,0)\,, \label{scalarfields}
	\eea 
	and
	\bea 
	\langle \eta \rangle   =  \left( \begin{array}{c} \fr {v_\eta}{\sqrt{2}} \\ 0 \\ 0\end{array} \right),  \, 
	\langle \chi \rangle   = \left( \begin{array}{c} 0 \\ 0\\ \fr {v_{\chi}}{\sqrt{2}}\end{array} \right),  \,
	\langle \rho \rangle  =  \left( \begin{array}{c} 0\\ \fr {v_\rho}{\sqrt{2}} \\ 0
	\end{array} \right),  \,
	\langle \phi \rangle  = \fr { v_\phi}{\sqrt{2}}\,.
	\label{eq2t}\eea
	
	As shown in Eq.(\ref{scalarfields}), $\eta$ and $\chi$ have the same quantum number under the act of $\mbox{SU(3)}_C\times \mbox{SU(3)}_L\times \mbox{U(1)}_X$ group, so in the self-coupling of scalars, non-Hermitian terms of order 2 such as $\eta^\dagger \chi$ or order 3 as $\eta^\dagger \chi \phi$, etc, certainly appear. These terms cause many difficulties in determining the physical state of Higgs bosons. Therefore, one can eliminate this difficulty by introducing the $Z_{11}$ into the gauge group \cite{Dias:2003iq}. There are several different ways to represent $Z_{11}$ to eliminate non-Hermitian terms, but in this work, we choose the same as Ref.\cite{alp331}. Then, the fields change under the act of $Z_{11}$ as follows: 
		
	\bea \label{Z11transform}
	Q_{3L}&\rightarrow& \omega_0Q_{3L},\crn
	\psi_{aL}&\rightarrow& \omega^{1}_1\psi_{aL},\hs \phi \rightarrow \omega^{-1}_1\phi,\crn
	d_{aR}&\rightarrow& \omega^{1}_2d_{aR},\hs \rho \rightarrow \omega^{-1}_2\rho,\crn
	\left(l_{aR},~U_{3R}\right) &\rightarrow& \omega^{1}_3\left(l_{aR},~U_{3R}\right),\hs \chi \rightarrow \omega^{-1}_3\chi,\crn
	D_{nR}&\rightarrow& \omega^{1}_4D_{nR},\hs Q_{nL} \rightarrow \omega^{-1}_4Q_{nL},\crn
	u_{aR}&\rightarrow& \omega^{1}_5u_{aR},\hs \eta \rightarrow \omega^{-1}_5\eta,
    \eea
    where $\om_m \equiv e^{ \fr{i 2\pi m}{11}}, m=0,\pm 1\cdots \pm 5$ are considered as charges of the $Z_{11}$ group.

    To generate masses for most SM fermions at the tree level, we use Yukawa interactions involving the scalar triplets $\eta,~\rho$ and its VEVs as given in Eq.(\ref{scalarfields},\ref{eq2t}). A most convenient approach is introduced where $\rho$ generates masses for charged leptons and d-quarks, and $\eta$ generates masses for u-quarks. In addition, exotic fermions such as $U_{3L,R},~D_{nL,R}$ can also acquire masses at the tree level through the VEV of $\chi$, gauge singlets $N_{aR}$ may be take masses through the VEV of $\phi$ field. The remaining fermions belong to the SM, which are active neutrinos that can acquire masses through a type 1 seesaw mechanism based on couplings with$N_{aR}$. To accomplish this, $Z_{2}$ is added to the gauge group to eliminate unwanted Yukawa interactions. On the other hand, $Z_{2}$ also contributes to PQ symmetry naturally to create a stable state of axion \cite{Dias:2003iq}. The $Z_{2}$ symmetry distinguishes particles in the model into two classes, even and odd, corresponding to charges of $1$ and $-1$. The odd fields of $Z_{2}$ group are given as \cite{alp331}, 
    	\be (\eta\,, \rho\,, u_{aR} \,,  d_{a R},
    	l_{aR}, N_{aR})\,\,\,\,\rightarrow \,\,\,\,
    	-\, (\eta\,, \rho\,,  u_{aR}\,,  d_{a R},
    	l_{aR}, N_{aR})\,.
    	\label{oct1}	\ee

	By introducing the discrete groups based on the above discussions, the full gauge symmetry group of the model is $\mbox{SU(3)}_C\otimes \mbox{SU(3)}_L\otimes \mbox{U(1)}_X \otimes Z_{11}\otimes Z_2$ and under the transformation of this  group the particle assignments are	summarized in table \ref{qnumber}. 
		\begin{table}[th]
			\scalebox{1.25}{\begin{tabular}{|c|c|c|c|c|c|c|c|c|c|c|c|c|c|c|}
					\hline
					& $Q_{n L}$ & $Q_{3L}$ & $u_{a R}$ & $d_{aR}$ & $U_{3R}$ &$ D_{nR} $ & $\psi_{aL}$ & $l_{aR}$
					& $N_{aR} $ & $\eta$ & $\chi $ & $\rho$ & $
					\phi$  \\ \hline
					$SU(3)_C$ & $\mathbf{3}$ & $\mathbf{3}$ & $\mathbf{3}$ & $\mathbf{3}$ & $%
					\mathbf{3}$ & $\mathbf{3}$ & $\mathbf{1}$ & $\mathbf{1}$ & $\mathbf{1}$ & $%
					\mathbf{1}$ & $\mathbf{1}$ & $\mathbf{1}$ & $\mathbf{1}$   \\ \hline
					$SU(3)_L$ & $\overline{\mathbf{3}}$ & $\mathbf{3}$ & $\mathbf{1}$ & $%
					\mathbf{1}$ & $\mathbf{1}$ & $\mathbf{1}$ & $\mathbf{3}$ & $\mathbf{1}$ & $%
					\mathbf{1}$ & $\mathbf{3}$ & $\mathbf{3}$ & $\mathbf{3}$ & $\mathbf{1}$  \\ \hline
					$U(1)_X$ & $0$ & $\fr 1 3$ & $\fr 2 3$ & $-\fr 1 3$ & $\fr{2
					}{3}$ & $-\fr 1 3$ & $-\fr 1 3$ & $-1$ & $0$ & $-
					\fr 1 3$ & $-\fr 1 3$ & $\fr 2 3$ & $0$
					\\ \hline
					$Z_{11}$ & $\om^{-1}_4$ & $\om_0$ & $\om_5$ & $\om_2$
					& $\om_3$ & $\om_4$ & $\om_1$ & $\om_3$ & $%
					\om^{-1}_5$ & $\om^{-1}_5$ & $\om^{-1}_3$ & $\om^{-1}_2$ & $\om^{-1}_1$  \\
					\hline
					$Z_2$ & $1$ & $1$ & $-1$ & $-1$ & $1$ & $1$ & $1$ & $-1$ &
					$-1 $ & $-1$ & $1$
					& $-1$ & $1$ \\ \hline
			\end{tabular}}
			\caption{The $SU(3)_C\otimes SU(3)_L\otimes U(1)_X\otimes Z_{11}\otimes Z_2$ charge assignments of spectrum of the particles in the model with $a=1,2,3$ and $n =1,2$.}
			\label{qnumber}
		\end{table}

		From table \ref{qnumber}, one can derive the Yukawa interactions of the model with two parts corresponding to quarks and leptons, in which the terms that do not preserve the charge of the symmetry gauge group are eliminated. The Yukawa couplings invariant under the $SU(3)_C\otimes SU(3)_L\otimes U(1)_X\otimes Z_{11}\otimes Z_2$ symmetry, arise \cite{alp331}:
		\bea
		-\mathcal{L}^{Y}_f&=&  -\mathcal{L}^{Y}_q	-\mathcal{L}^{Y}_\ell, \crn
		-\mathcal{L}^{Y}_q&=& y_{1}\bar{Q}_{3L}U_{3R}\chi +\sum\limits_{n,p =1}^{2}\left(	y_{2}\right) _{np }\bar{Q}_{n L}D_{p R}\chi ^{\ast}\crn
		&&+\sum\limits_{i=1}^{3}\left( y_{3}\right) _{3a}\bar{Q}_{3L}u_{aR}\eta
		+\sum\limits_{n =1}^{2}\sum\limits_{i=1}^{3}\left( y_{4}\right) _{ni}\bar{Q}_{n L}d_{iR}\eta ^{\ast }\crn
		&&+\sum\limits_{i=1}^{3}\left( y_{5}\right) _{3i}\bar{Q}_{3L}d_{iR}\rho
		+\sum\limits_{n =1}^{2}\sum\limits_{i=1}^{3}\left( y_{6}\right) _{ni}\bar{Q}_{n L}u_{iR}\rho ^{\ast },\label{yukinterac} \\
		-\mathcal{L}^{Y}_\ell &=&\sum\limits_{i=1}^{3}\sum\limits_{j=1}^{3}g_{ij}%
		\bar{\psi}_{iL}l_{jR}
		\rho	+\sum\limits_{i=1}^{3}\sum\limits_{j=1}^{3}\left( y_{\nu }^{D}\right) _{ij}%
		\bar{\psi}_{iL}\eta N_{jR}\crn
		&&+\sum\limits_{i=1}^{3}\sum\limits_{j=1}^{3}\left(
		y_{N}\right) _{ij}\phi \bar{N}_{iR}^{C}N_{jR}+\mbox{H.c.}
		\label{yukintera}
		\eea
		We can easily see that the Lagrangian in Eqs.(\ref{yukinterac},~\ref{yukintera}) is conserved charge for all elements of the gauge group, including the $Z_2$ group as usual. It is emphasized that the transformation under the $Z_2$ as in Eq. (\ref{oct1}) is different from the one given in Ref. \cite{Ferreira:2016uao} where $\chi,~D_{nR},~U_{3R}$ are odd. Of course, choosing the act of $Z_2$ as Ref.\cite{Ferreira:2016uao} or Eq.(\ref{oct1}) does not change Yukawa interactions in Eq.(\ref{yukinterac},~\ref{yukintera}). Accordingly, $\chi$ only couple with exotic quarks and is responsible for giving mass to particles outside the standard model except $N_{aR}$. The masses of $N_{aR}$ are generated through $v_\phi$ of the gauge singlet $\phi$. Active neutrinos acquire masses via a type 1 seesaw mechanism based on the 2nd and 3rd terms of Eq.(\ref{yukinterac}). The mass matrix in the basis ($\nu_{aR},~N_{aR}$) is,
		\bea 
		\mbox{M}^{\nu}=\left( \begin{array}{cc}
			\mathbf{O}_{3\times3} & M_\nu^D \\
			\left(M_\nu^D \right)^T& M_N\\
			\end{array}\right),
		\eea
		where $M_\nu^D=\frac{y^D_\nu v_\eta}{\sqrt{2}},~ M_N=\sqrt{2}y_N v_\phi$ are $3\times3$ matrices and the masses of active neutrinos are given based on the type 1 seesaw mechanism as \cite{Ibarra:2010xw},
		\bea \label{type1ss}
		m_\nu=M_\nu^DM^{-1}_N\left(M_\nu^D \right)^T.
		\eea
		
		Next, let's consider a role of VEVs of scalars in creating masses for particles in the model. Firstly, the VEV of $v_{\phi}$ breaks the $Z_{11}$ symmetry to helps the axion-like particle and an inflation arise. Secondly, VEV of $\chi$ breaks the flavor symmetry $SU(3)_L$ into $SU(2)_L$ group and generates massess for the exotic fermions. Thirdly, two remaining $SU(3)_L$ triplet scalars $\eta$ and $\rho$ generate masses for SM particles  as in the usual 3-3-1 models. To fully illustrate this, we present the following scheme of symmetry breaking:
		
		\begin{eqnarray} \label{scheme}
			SU(3)_{C}\otimes SU(3)_{L} &\otimes &U(1)_{X}\otimes Z(11)\otimes Z_{2} 
			\notag \\
			&\downarrow &v_{\phi }  \notag \\
			SU(3)_{C}\otimes SU(3)_{L} &\otimes &U(1)_{X}\otimes Z_{2}
			\crn
			&\downarrow &v_{\chi }  \notag \\
			SU(3)_{C}\otimes SU(2)_{L} &\otimes &U(1)_{Y}\otimes Z_{2}
			\crn
			&\downarrow &v_{\eta},\, v_{\rho} \notag \\
			SU(3)_{C}&\otimes &U(1)_{Q} \otimes Z_{2}
		\end{eqnarray}%
		Clearly, $Z_2$ is a residual symmetry after the spontaneous symmetry-breaking stages, and it plays a crucial role in identifying stable particles in the model, which may include dark matter candidates.
		\subsection{Gauge bosons}
		We consider structure of gauge bosons from the nine gauge fields of the electroweak group $SU(3)_L\times U(1)_X$ and the masses of these bosons derived from VEVs of scalars. They all originate from the kinetic energy of scalars, with the familiar form given as, 
		\be \mathcal{L}_{S} = \sum_{S=\chi, \eta,\rho,\phi} (D^\mu S)^\dag D_\mu S \,,
		\label{j231}
		\ee
		with $D_{\mu}$ is covariant derivative and given as follow,
		\be  D_{\mu}\equiv \pa _{\mu}-i g T^a W^a_{\mu}-i g_X X T^9 B_{\mu} \equiv \pa_\mu - i P_\mu ,  \label{j232}\ee
		where  $T^9=\frac{I_{3\times 3}}{\sqrt{6}}$ with $I_{3\times 3}$ is the $3\times 3$ identity matrix,  $W^a_{\mu}$ ($a=\overline{1,8}$) are gauge fields of $SU(3)_L$, $B_{\mu}$ is gauge field of $U(1)_X$,  and $g$, $g_X$ are gauge couplings of the two groups $SU(3)_L$ and $U(1)_X$, respectively. Furthermore, the matrix $W^aT^a$, where $T^a = \frac{\la_a}{2}$ corresponds to a triplet representation, is written as follows:
		\bea P_\mu  =\frac{g}{2}\left(
		\begin{array}{ccc}
			W^3_{\mu}+\fr{1}{\sqrt{3}} W^8_{\mu} + t\sqrt{\fr 2 3} X B_\mu & \sqrt{2}W^+_{\mu} &  \sqrt{2} X^{0}_{\mu} \\
			\sqrt{2}W^-_{\mu} &  -W^3_{\mu}+\fr{1}{\sqrt{3}} W^8_{\mu}+  t\sqrt{\fr 2 3} X B_\mu & \sqrt{2}Y^{-}_{\mu} \\
			\sqrt{2}X^{0*}_{\mu}& \sqrt{2}Y^{+}_{\mu} &-\fr{2}{\sqrt{3}} W^8_{\mu}+ t\sqrt{\fr 2 3} X B_{\mu} \\
		\end{array}
		\right),
		\label{j233}\eea
		in which we have denoted $t = \frac{g_X}{g}$ and  defined the mass eigenstates of the charged gauge bosons as:
		\bea W^{\pm}_{\mu } &= &\fr{1}{\sqrt{2}}\left( W^1_{\mu}\mp i W^2_{\mu}\right),\hs 
		Y^{\pm }_{\mu} = \fr{1}{\sqrt{2}}\left( W^6_{\mu}\pm i W^7_{\mu}\right),\crn 
		X^{ 0}_{\mu} & = & \fr{1}{\sqrt{2}}\left( W^4_{\mu}- i W^5_{\mu}\right),\hs
		X^{0*}_{\mu}=\fr{1}{\sqrt{2}}\left( W^4_{\mu}+i W^5_{\mu}\right) .
		\label{j234}\eea
		
		After spontaneous  symmetry breaking (SSB), the mass spectrum of the gauge bosons arises from the following kinertic terms, 
		\be 
		\mathcal{L}_{mass} = \sum_{S=\chi,\eta,\rho}(D^\mu\langle S \rangle)^\dag (D_\mu \langle S\rangle)\,.
		\ee
		The $W$ boson which is identical to that in the SM and its mass is given as,
		\begin{equation}
			m^2_{W} = \fr{g^2}{4}(v_\eta^2+v_\rho^2), \label{Wmass}
		\end{equation}
		while two $bilepton$ gauge bosons $(X,Y)$ get masses as below \cite{il}:
		\begin{equation}
		 m^2_{X^0}=\fr{g^2}{4}(v_{\chi}^2+v_\eta^2),\hs m^2_Y=\fr{g^2}{4}(v_{\chi}^2+v_\rho^2)\, \label{YXmass}
		\end{equation}
		and their masses splitting with $W$ boson is \cite{il}
		\[|m^2_Y-m^2_{X^0}|< m^2_W\,.\]
		
		From Eq.\eq{Wmass}, it follows
		\be v_\eta^2+v_\rho^2 = v^2 = 246^2 \,  \textrm{GeV}^2\,.
		\label{j236} \ee
		In the basis $(W^3_\mu, W^8_\mu, B_\mu)$, the squared mass mixing matrix of neutral gauge bosons is
		\bea
		\mathcal{M}^2_{NG}=\left(
		\begin{array}{ccc}
			\frac{g^2}{4}(v_\eta^2 + v_\rho^2) & \frac{g^2}{4\sqrt3}(v_\eta^2 - v_\rho^2)  &  -\frac{gg_X}{6\sqrt6}(v_\eta^2 + 2v_\rho^2)\\
			&\frac{g^2}{12}(v_\eta^2 + v_\rho^2 + 4 v_\chi^2)  &  \frac{gg_X}{18\sqrt2}(2 v_\rho^2 + 2 v_\chi^2 - v_\eta^2)\\
			&  & \frac{g_X^2}{54}(v_\eta^2 + 4 v_\rho^2 +  v_\chi^2)
		\end{array}\right)\,.\label{MassNeutralGauge}
		\eea
		The matrix $\mathcal{M}^2_{NG}$ has three eigenvalues which belong to  photon $A$ and $Z$, $Z^\prime$ bosons, respectively:
		\bea \label{mngauge}
		m^2_A &=& 0\,, \notag\\
		m^2_Z &=& \frac{g^2(9g^2 + 2 g_X^2)(v_\eta^2 v_\rho^2 + v_\rho^2 v_\chi^2 + v_\chi^2 v_\eta^2)}{2\left(18g^2(v_\eta^2 + v_\rho^2 +v_\chi^2)+g_X^2 (v_\eta^2 + 4v_\rho^2 +v_\chi^2)\right)}\,, \notag\\
		m^2_{Z^\prime}&=& \frac{g^2}{3} (v_\eta^2 + v_\rho^2 +v_\chi^2) + \frac{g_X^2}{54}(v_\eta^2 + 4v_\rho^2 +v_\chi^2) - m^2_Z\,.\eea
		
		The physical states of these neutral gauge boson are defined as below:
		\bea
		\left(\begin{array}{c}
			A_\mu \\ Z_\mu \\ Z^\prime_\mu 
		\end{array} \right) = \left(
		\begin{array}{ccc}
			c_{\beta_2} & c_{ \beta_1} s_{ \beta_2} & s_{ \beta_1} s_{ \beta_2} \\
			-s_{\beta_2} & c_{ \beta_1} c_{ \beta_2} & c_{ \beta_2} s_{ \beta_1} \\
			0 & -s_{ \beta_1} & c_{\beta_1} \\
		\end{array}
		\right)\left( \begin{array}{c}
			W^3_\mu \\ W^8_\mu \\ B_\mu 
		\end{array}\right)\,,
		\eea
		where, we denote $c_\varphi = \cos \varphi$, $s_\varphi = \sin \varphi$, $t_\varphi = \tan \varphi$ and the mixing angles are given by:
		\bea \label{betangle}
		t_{ 2 \beta_1} &=& \frac{6\sqrt2 t_{W} (v_\eta^2 -2v^2_\rho -2v^2_\chi)}{2 t_{W}^2 (v^2_\eta +4 v^2_\rho + v^2_\chi)-9(v^2_\eta + v^2_\rho +4 v^2_\chi)}\,, \notag \\
		t_{ 2 \beta_2} &=&
		-\frac{3 \sqrt{6} c_W \left(B \left(v_\eta^2-v_\rho^2\right)+2 \tan ^2\theta_W \left(v_\eta^4 - 3 v_\eta^2 v_\rho^2 -5v^2_\eta v_\chi^2-7 v_\rho^2 v_\chi^2-4 v_\rho^4\right)\right)}{\left(v_\eta^2-2 v_\rho^2-2v_\chi^2\right) \left(A-2 t ^2_W \left(v_\eta^2+4 v_\rho^2+v_\chi^2\right)+45 \left(v_\eta^2+v_\rho^2\right)-36 v_\chi^2\right) \sqrt{1+\frac{\left(B ct_W-2 t_W \left(v_\eta^2+4 v_\rho^2+v_\chi^2\right)\right)^2}{72 \left(v_\eta^2-2 v_\rho^2-v_\chi^2\right)^2}}}\,, \notag\\
		\eea
		with $s_w\equiv \sin\theta_W $,~ $\theta_W$ is Weinberg angle as known, and 
		\bea
		A &=& \sqrt{\left(2 t^2_W \left(v_\eta^2+4 v_\rho^2+v_\chi^2\right)+9   \left(v_\eta^2+v_\rho^2+4 v_\chi^2\right)\right)^2-18 (v_\eta^2 v_\rho^2+ v_\eta^2 v_\chi^2+ v_\rho^2 v_\chi^2)}\,,\notag\\
		B &=& 9  \left( v_\eta^2+v_\rho^2+4 v_\chi^2\right) -A \,.
		\eea
		In the limits $v_\chi \gg v_\eta, v_\rho$, the mixing angle $\beta_2$ is very tiny, approximates zero, while the mixing angle $\beta_1$ is redefined as:
		\be t_{ 2\beta_1} \approx \frac{6\sqrt2 t_W}{t^2_W -18}\,.
		\ee
		\subsection{The scalar sector}
		We recall that the self-couplings of the scalars can have non-Hermitian terms such as: $\eta^\dagger\chi$, $\chi^\dagger\eta\phi^*$, $(\chi^\dagger\eta)^2$ because $\eta$ and $\chi$ have the same quantum numbers as given in Eq.(\ref{scalarfields}). However, we have the selections as shown in Refs.\cite{Dias:2003iq,alp331} to remove the non-Hermitian terms by adding $Z_{11}$ in the gauge group. Next, we just need to add $Z_2$ group to automatically perform the PQ symmetry ensuring the naturally stable and light of the axion. The acts of $Z_{11}$ and $Z_2$ symmetry in this work are the same as 331ALP, which was mentioned in Ref.\cite{alp331}. The Higgs potential only has Hermitian terms as follows: \cite{alp331}:
		\bea
		V &=& \mu^2_\phi  \phi^* \phi +  \mu_\chi^2  \chi^\dag \chi + \mu_\rho^2
		\rho^\dag \rho +  \mu_\eta^2 \eta^\dag \eta + \la_1 ( \chi^\dag \chi)^2 + \la_2 ( \eta^\dag
		\eta)^2\crn &  & + \la_3 ( \rho^\dag \rho)^2 +
		\la_4 ( \chi^\dag \chi)( \eta^\dag \eta) + \la_5 ( \chi^\dag \chi)( \rho^\dag \rho) + \la_6 ( \eta^\dag \eta)( \rho^\dag \rho)\crn
		&& + \la_7 ( \chi^\dag \eta)( \eta^\dag \chi) + \la_8 ( \chi^\dag \rho)( \rho^\dag \chi) + \la_9 ( \eta^\dag \rho)( \rho^\dag \eta)\crn
		&& + \la_{10} ( \phi^* \phi)^2 + \la_{11} ( \phi^* \phi)( \chi^\dag \chi) + \la_{12} ( \phi^* \phi)( \rho^\dag \rho)\crn
		&& + \la_{13} ( \phi^* \phi)( \eta^\dag \eta)
		+ \left( \la_\phi\ep^{ijk} \eta_i \rho_j \chi_k \phi
		+H.c.\right). \label{poten3} \eea
		
		These scalar fields should be expanded around their VEVs, 
		\bea
		\rho_2^0 &= & \fr{1}{\sqrt{2}}(v_\rho + R^2_\rho + i I^2_\rho)\, , \hs \eta_1^0 = \fr{1}{\sqrt{2}}(v_\eta + R^1_\eta + i I^1_\eta)\, ,\crn
		\chi_3^0 &= & \fr{1}{\sqrt{2}}(v_\chi + R^3_\chi + i I^3_\chi)\, , \hs \phi  = \fr{1}{\sqrt{2}}(v_\phi + R_\phi + i I_\phi)\, . \label{poten4}
		\eea
		
		Substituting (\ref{poten4})  into  (\ref{poten3})  leads to the following minimum potential conditions,
		\bea
		\mu^2_\rho + \la_3 v_\rho^2 + \fr{\la_5}2 v_\chi^2 + \fr{\la_6}2 v_\eta^2 + \fr{\la_{12}}2 v_\phi^2 + \fr{A}{2 v_\rho^2} & = & 0\, ,\crn
		\mu^2_\eta + \la_2 v_\eta^2 + \fr{\la_4}2 v_\chi^2 + \fr{\la_6}2 v_\rho^2 + \fr{\la_{13}}2 v_\phi^2 + \fr{A}{2 v_\eta^2} & = & 0\, ,\crn
		\mu^2_\chi + \la_1 v_\chi^2 + \fr{\la_4}2 v_\eta^2 + \fr{\la_5}2 v_\rho^2 + \fr{\la_{11}}2 v_\phi^2 + \fr{A}{2 v_\chi^2} & = & 0\, ,\crn
		\mu^2_\phi + \la_{10}  v_\phi^2 + \fr{\la_{11}}2 v_\chi^2 + \fr{\la_{12}}2 v_\rho^2 + \fr{\la_{13}}2 v_\eta^2 + \fr{A}{2 v_\phi^2} & = & 0\, , \label{poten5}
		\eea
		where $\mathcal{A} \equiv \la_\phi v_\phi v_\chi v_\eta  v_\rho $.

		The Higgs potential in Eq.(\ref{poten3}) gives the mass mixing matrix of the charged scalar fields, diagonalizing it allows to give the physical state and masses of the charged Higgs bosons. The result obtained is similar to Ref.\cite{alp331},
		\be m^2_{H^{\pm}_1}
		=- \fr{v^2}{2}\left(\fr{\mathcal{A}}{v_\rho^2 v_\eta^2}-\la_9\right)\,,~  m^2_{H^{\pm}_2} =-\fr{(v_\chi^2 + v_\rho^2)}{2}\left(\fr{\mathcal{A}}{v_\chi^2 v_\rho^2}-\la_8\right)\,,
		\label{massch}
		\ee
		with conditions $\la_9 > \la_\phi \fr{v_\phi v_\chi}{v_\rho v_\eta}$, ~$\la_8 > \la_\phi \fr{v_\phi v_\eta}{v_\chi v_\rho}$  and 
		\be
		\left(
		\begin{array}{c}
			G_W^{\pm} \\
			H^{\pm}_{1} \\
		\end{array}
		\right) = \left(
		\begin{array}{cc}
			c_ \al  & - s_ \al  \\
			s_ \al  &  c_\al  \\
		\end{array}
		\right) \left(  \begin{array}{c}
			\rho_1^{\pm}\\
			\eta_2^{\pm} \\
		\end{array}
		\right)\, ,~\left(
		\begin{array}{c}
			G_Y^{\pm} \\
			H^{\pm}_2 \\
		\end{array}
		\right) = \left(
		\begin{array}{cc}
			c_{ \theta_1}  & - s_{ \theta_1 }\\
			s_{ \theta_1} &  c_{ \theta_1}   \\
		\end{array}
		\right) \left(  \begin{array}{c}
			\chi_2^{\pm} \\
			\rho_3^{\pm}\\
		\end{array}
		\right)\, ,
		\label{statech}
		\ee
		where the mixing angles are defined by $t_\al  =\fr{v_\eta}{ v_\rho}$,~$t_{\theta_1}=\fr{v_\rho}{ v_\chi}$, and $G_W^{\pm}$,~$G_Y^{\pm}$ are the Goldstone bosons associated with the longitudinal component of the $W^{\pm}$,~ $Y^{\pm}$, respectively.
		
		Next, we are interested in the mass matrix of the  CP-odd scalars. According to Eq.(\ref{scalarfields}), we have six initial pseudoscalars ($I^3_\eta,~I^1_\chi,~I_\phi,~I^3_\chi,~I^2_\rho,~I^1_\eta$). However, we can separate them into two uncouplings bases \cite{Dias:2003iq} depending on the VEVs of scalars (\ref{poten4}) and the Higgs potential (\ref{poten3}). The first base is ($I^3_\eta,~I^1_\chi$), corresponding to a $2\times2$ mass matrix, and the second base is ($I_\phi,~I^3_\chi,~I^2_\rho,~I^1_\eta$), corresponding to a $4\times4$ mass matrix. The mass matrix in the first base is easily diagonalized to give two physical states, but they are ignored due to their lack of significant relationship. In this work, we are interested in the $4\times4$ mass matrix in the ($I_\phi,~I^3_\chi,~I^2_\rho,~I^1_\eta$) base, whose form is \cite{alp331},
		\bea
		{M_{I}^2= - \fr {\mathcal{A}}{2}   \left(
			\begin{array}{cccc}
				\fr 1 {v_\phi^2} & \fr 1 {v_\phi v_{\chi}}&  \fr 1 {v_\phi v_\rho} &  \fr 1 {v_\phi v_\eta}  \\
				& \fr 1 {v_{\chi}^2}&  \fr 1 {v_{\chi} v_\rho} &  \fr 1 {v_{\chi} v_\eta}  \\
				&  & \fr 1 {v_\rho^2} &  \fr 1 {v_\eta v_\rho} \\
				&  &  & \fr 1 {v_\eta^2} \\
			\end{array}
			\right)\, .
			\label{Modd5}}
		\eea
		
		We can use the Euler method to exactly diagonalized mass matrix in Eq.(\ref{Modd5}) as shown in Ref.\cite{alp331}. The results obtained include: two massless states, namely $G_Z$ and $G_{Z^\prime}$ absorbed by $Z$ boson and $Z^\prime$ boson, a massive pseudoscalar and an axion. With a similar context as shown in Refs.\cite{Dias:2003iq, Dias:2002hz}, the model under consideration also exists a automatic PQ symmetry in classical Lagrangians, where we can eliminate trilinear terms in the Higgs potential. Fortunately, we can accomplish this by adding $Z_2$ to the gauge group as mentioned above. To obtain a stability axion, we add $Z_{11}$ to the gauge group to prevent the PQ symmetry from being broken by gravity effects. Interestingly, both $Z_{11}$ and PQ symmetry are automatically implemented, resulting in a naturally light axion in the model \cite{Dias:2002hz}. However, there is a difference from Refs.\cite{Dias:2003iq,Ferreira:2016uao} that the axion state obtained in this work is a combination of ($I_\phi,~I^3_\chi,~I^2_\rho,~I^1_\eta$). The relationships between CP-odd scalars in the physical states and the interactive states are represented as below: 
		\bea
		a &=& I_\phi c_{\theta_\phi} - I_\chi^3 s_{\theta_3} s_{\theta_\phi} - I_\rho^2 s_
		\al c_{ \theta_3} s_{ \theta_\phi} - I_\eta^1 c_ \al
		c_{\theta_3} s_{ \theta_\phi}\,, \label{ALPphys}\\
		G_{Z^\prime} &=& I_\chi^3 c_{ \theta_3} - I_\rho^2 s_\al s_{\theta_3} - I_\eta^1  c_
		\al s_{\theta_3}\,,\\
		G_Z &=& I_\rho^2 c_ \al - I_\eta^1 s_\al\,,\\
		A_5 &=& I_\phi s_{ \theta_\phi} + I_\chi^3 s_{ \theta_3} c_{ \theta_\phi} + I_\rho^2 s_ \alpha c_{ \theta_3} c_{\theta_\phi} + I_\eta^1 c_ \al c_{\theta_3} c_{ \theta_\phi}\,, \label{physstates}
		\eea
		with $\al$ is given above, two remain angles in this sector are:
		\bea
		t_{ \theta_3} 
		= \fr{v_\eta}{v_{\chi}} |c_ \alpha|\,, \hs t_{ \theta_\phi}
		= \frac{1}{v_\phi \sqrt{\fr 1 {v_\chi^2} + \fr 1 {v_\rho^2} +\fr 1 {v_\eta^2}}} 
		\,. \label{mixodd}
		\eea
		Because of the VEVs' hierachy $v_\rho, v_\eta \ll v_\chi \ll v_\phi$ as initial assumption, the derived mixing matrix in Eq.\eq{physstates} has just three angles $\al, \theta_3, \theta_\phi$ given and one parameter $ \left( \fr 1{v^2_\phi} + \fr 1{v^2_{\chi}} + \fr 1{v^2_\rho} +\fr 1{v^2_\eta}  \right)$ which is contained in the expression of $A_5$ mass  in Eq.\eq{mA5}. Furthermore, the mass of new massive field CP-odd scalar field ${A_5}$ is given by:
		\bea m^2_{A_5}  & = &  -\fr{\mathcal{A}}{2}  \left( \fr 1{v^2_\phi} + \fr 1{v^2_{\chi}} + \fr 1{v^2_\rho} +\fr 1{v^2_\eta}  \right)\,
		\approx -\fr{\la_\phi v_\phi v_\chi}{s_{ 2\al}}
		\, .\label{mA5}
		\eea
		The Eq. \eq{mA5}  shows that the values of $\la_\phi$ should be negative. It is emphasized that  both of the squared  mass mixing matrix in Eq. \eq{Modd5} and 		mass of the $A_5$ field are arised from the last term in Eq.\eq{poten3} which just  appears because of the	specific discrete symmetry $Z_{11} \otimes Z_2$ imposing to this model from the beginning \cite{Dias:2003iq}. Here, the axionlike particle $a$ is massless and given by the combination of  four CP odd neutral scalar fields  $I_\phi$, $I_{\chi}^3$, $I_\rho^2$ and $I_\eta^1$ as in Eq.(\ref{ALPphys}). Due to 	$v_{\chi} \ll v_\phi$, the values of $t_{ \theta_\phi}$ as well as $s_{ \theta_\phi}$ go to zero, then $c_{\theta_\phi} \backsimeq 1$ which approximately helps $a \backsimeq I_\phi$.
		
		 Similarly to the pseudoscalars part, we also have six original CP-even scalars divided into two uncoupling bases. Those two bases are ($R^3_\eta,~ R_{\chi}^1$) and ($R_\phi,~ R_{\chi}^3,~ R_\rho^2,~ R_\eta^1$). The mass matrix in ($R^3_\eta,~ R_{\chi}^1$) base is easily diagonalized to yield  masses of particles because it is a $2\times 2$ matrix. The mass matrix in the remaining base is $4\times 4$ and difficult to diagonalize because it has nontrivial eigenvalues. We can use the Hartree-Fock method to diagonalize it as shown in Ref.\cite{alp331}. In this way, the original $4\times4$ matrix is diagonalized in decreasing order through three unitary matrices. Three rotation angles characteristic of those matrices will be used as dependent parameters in numerical investigation below. We return to the mass matrix in the $(R^1_\eta, R_\rho^2, R_\chi^3, R_\phi)$ base, its form is \cite{alp331},   
		
		\bea
		M_R^2 =       {2} \left(
		\begin{array}{cccc}
			\la_2 v_\eta^2-\fr{\mathcal{A}}{4 v_\eta^2} &
			\fr{\la _6 v_\eta v_\rho}{4} +\fr{\la_\phi  v_\chi v_\phi}{4} & \fr{\la_4 v_\eta
				v_\chi}{4} +\fr{\la_\phi  v_\rho
				v_\phi}{4} & \fr{\la_\phi
				v_\rho v_\chi}{4}+\fr{\la _{13}
				v_\eta v_\phi}{2}  \\
			& \la_3 v_\rho^2-\fr{\mathcal{A}}{4
				v_\rho^2} & \fr{\la_\phi
				v_\eta v_\phi}{4} +\fr{\la _5 v_\rho v_\chi}{2} &  \fr{\la
				\phi  v_\eta v_\chi}{4}+\fr{\lambda _{12}
				v_\rho v_\phi}{2} \\
			& & \la_1 v_\chi^2-\fr{\mathcal{A}}{4 v_\chi^2} & 
			\fr{\la\phi  v_\eta v_\rho}{4}+ \fr{\la_{11} v_\chi v_\phi}{2}
			\\
			&  & 
			& \la_{10} v_\phi^2-\fr{\mathcal{A}}{4 v_\phi^2} \\
		\end{array}
		\right)\,. \label{MR}
		\eea
		Three rotation angles given to diagonalize the matrix $M^2_R$ in Eq.(\ref{MR}) using the Hartree-Fock method are, 
		\bea \label{ta2}
		&& t_{ 2 \al_2} = \fr{4 c_ {\al_3} v_\eta v_\rho (\mathcal{A} + \la_6 v_\eta^2 v_\rho^2)}{\mathcal{A} c^2_{\al_3}v_\eta^2 -\mathcal{A} v_\rho^2 + 4 v_\eta^2 v_\rho^2 (\la_2 v_\eta^2 - \la_3 c^2_{\al_3} v_\rho^2)}\,, \\
		&&  t_ {2\al_3 }= \fr{4 v_\chi
			\left(\mathcal{A}+2 \la_5 v_\rho^2
			v_\chi^2\right)}{ c_ {\al_\phi}  \left(\mathcal{A}-4 \la_1	v_\chi^4\right)^2}\,, \\
		&& t_{ 2 \al_\phi} = \fr{\la_{11} v_{\chi }}{\la_{10} v_{\phi }} \,.
		\eea
		
		After diagonalizing the $M^2_R$ matrix, four physical states of CP-even Higgs are given. Two particles have masses in electroweak scale ($h_1, h_2$), with the lighter particle ($h_1$) being identified as the SM-like Higgs boson. The remaining two particles are thought to be heavier, one proportional to $v_\chi$ and the other in the $v_\phi$ scale. These physical states are: 
		\bea
		h_1 &=& R_\eta^1 c_{ \al_2} + R^2_\rho s_{ \al_2} c_{ \al_3} + R_\chi^3 s_{ \al_2}s_{ \al_3} c_{ \al_\phi} - R_\phi s_{ \al_2}  s_{ \al_3} s_{ \al_\phi}\,,\\
		h_2 &=& -R_\eta^1 s_{ \al_2} + R^2_\rho c_{ \al_2} c_{ \al_3} + R_\chi^3 c_{ \al_2}s_{\al_3} c_{ \al_\phi} - R_\phi c_{ \al_2}  s_{ \al_3} s_{ \al_\phi}\,,\\
		H_\chi &=& - R^2_\rho s_{ \al_3} + R_\chi^3 c_{ \al_3} c_{ \al_\phi} - R_\phi c_{ \al_3} s_{ \al_\phi}\,,\\
		\Phi &=& R_\chi^3 s_{ \al_\phi} + R_\phi c_{ \al_\phi}\,.
		\label{PhysCPeven}		\eea
		
		In the limit $v_\phi \gg v_\chi \gg v_\rho, v_\eta$, these physical fields have  approximate states and respective masses as below: 
		\bea
		h_1 &\approx& R^1_\eta c_{\al_2}+ R^2_\rho s_{ \al_2} \,, \mbox{~with~}m_{h_1,h_2}^2 \approx\lambda_3 v^2+\fr{m_{A_5}^2}{2 } \mp \sqrt{m_{A_5}^4 +\lambda_3^2 \left(v^4-3v_\eta^2 v_\rho^2\right)-\fr{\lambda_3 m_{A_5}^2
				\left(v^4-2v_\eta^2 v_\rho^2\right)}{v^2}}\,, \crn
		h_2 &\approx& -R^1_\eta s_{ \al_2}+ R^2_\rho c_{ \al_2}  \,, \crn
		H_\chi &\approx& R^3_\chi   c_{ \al_\phi}\,, \mbox{~with~} m_{H_{\chi}}^2   {\approx 2 \lambda_1 v_\chi^2 + \fr{\lambda_5^2}{2 \lambda_1} v_\rho^2}\,,\crn
		\Phi &\approx& R_\phi c_{ \al_\phi} \,,\mbox{~with~}  m_\Phi = \sqrt{2  \lambda_{10}v_\phi}.\,\label{physmassstate}
		\eea 	
		In comparision with the $4 \times 4$ mass mixing matrix of CP-odd sector containing only four parameters with three massless solution,  the matrix in Eq.\eq{MR} having ten parameters which is unable to be exactly diagonalized. To solve this problem we have used the Hartree-Fock method with 
		conditions such as $v_\phi \gg v_\chi \gg v_\rho, v_\eta$, $\la_\phi \ll 1$  and $s_{\al_3} \approx 0$.
		
		With these VEV hierarchy, the derived matrix contains three angles $\al_2,\al_3$ and $\al_\phi$ and three parameters associated  with masses of new fields $\Phi,~ H_\chi$ and $h_2$. The $H_\chi$ field is the heavy scalar boson which mass should be at TeV scale. The $\Phi$ field with heavy mass ranging about $v_\phi$ might be used to explain the inflation of the Early Universe \cite{ alp331, Long:2024rdq}. We can estimate $v_\phi$ by keeping the PQ symmetry preserved against the effects of quantum gravity by adding $Z_{11}$ group. Indeed, $Z_{11}$ can eliminate the effective operators (of the form $\frac{\phi^{N-1}}{M_{Pl}^{N-5}}$ for $Z_N, ~ N=11$) and makes PQ symmetry an automatic symmetry of classical Lagrangians. We request the Peccei-Quinn solution to the $\theta$ - strong problem is warranted, it mean that 			$\theta \sim \frac{v_\phi^N}{M_{Pl}^{N-4}\Lambda^4_{QCD}} \simeq 10^{-11}$. We recall that the instanton induced mass of axion is given as $m_a \sim \frac{\Lambda^2_{QCD}}{v_\phi}$ and $M_{Pl}$ is assumed at $10^{19}~\mathrm{GeV}$ \cite{Dias:2002hz, Dias:2003zt,Dias:2002gg}, so $v_\phi \sim 10^{11}~\mathrm{GeV}$ when $N=11$. Unfortunately, $\Phi,~ H_\chi$ fields are almost devoid of lepton flavor violating interactions, based on Eqs.(\ref{yukintera}, \ref{ta2}, \ref{physmassstate}). Only two lighter Higgs bosons, named as $h_1,~h_2$, are involved in lepton flavor violating processes, which we will consider below. Both of these particles have masses at electroweak scale, and the lighter particle, $h_1$,  is identified with the SM-like Higgs boson. 
		
		\section{Lepton flavor violating decay of charged lepton and CP-even Higgs bosons }
		\label{hLFV}
		In this section, we will be interested in the flavor lepton number violating decays $(l_a \rightarrow l_b\gamma)$ and $(H \rightarrow l_al_b),\, H\equiv h_1,\,h_2$. These channels are closely related, because they share a common source of lepton-flavor violation. Based on known experimental constraints, we will point out suitable regions of the model parameter space where interesting phenomenology and new physics may be found.
		\subsection{Couplings and Analytical form related  to LFV processes}
		\label{couplings}
		All lepton flavor violating interactions can be given based on the Lagrangian mentioned in Eq.(\ref{yukintera}), we rewrite them in a convenient form as follows:
		\bea
		-\mathcal{L}^{Y}_{Lepton}&=&g_{ab}\overline{\psi}_{aL}l_{bR}\rho
		+\left( y_{\nu }^{D}\right) _{ab}\overline{\psi}_{aL}\eta N_{bR}+\left(
		y_{N}\right) _{ab}\phi \overline{N}_{aR}^{C}N_{bR}+\mbox{H.c.},\crn
		&=& g_{ab}\left(\overline{\nu}_{aL}\rho^+_1+\overline{l}_{aL}\rho^0_2+\overline{(\nu_R^c)}_{aL}\rho^+_3\right)l_{bR}+ \left( y_{N}\right) _{ab}\phi \overline{N}_{aR}^{C}N_{bR}\crn 
		&+&\left( y_{\nu }^{D}\right) _{ab}\left( \overline{\nu}_{aL}\eta_1^0 + \overline{l}_{aL}\eta_2^- +\overline{\nu_R^c}_{aL}\eta_3^0 \right)  N_{bR} + \mbox{H.c.}
		\label{yuk}
		\eea
		The mass mixing matrices of the leptons are given:
		\bea
		-\mathcal{L}^{mass}_{Lepton}=\frac{g_{ab}v_\rho}{\sqrt{2}}\overline{l}_{aL}l_{bR}+ \frac{ \left( y_{N}\right)  _{ab}v_\phi}{\sqrt{2}} \overline{N}_{aR}^{C}N_{bR}
		+ \frac{ \left( y_{\nu }^{D}\right) _{ab}v_\eta}{\sqrt{2}}\overline{\nu}_{aL} N_{bR}  + \mbox{H.c.},
		\label{yuk1}
		\eea
		with the corresponding assignment,
		\be \label{ssmatrix}
		m_{l_a}=\frac{g_{ab}v_\rho}{\sqrt{2}}\delta_{ab},\hs  M^D_{\nu}=y^{D}_{\nu}\fr{v_{\eta}}{\sqrt{2}},
		\hs  M_N=\sqrt{2}\,y_{N}v_{\phi},
		\ee
		and
		\be
		g_{ab}=\frac{\sqrt{2}m_{l_a}}{v_\rho}\delta_{ab},\hs  y^{D}_{\nu}=\fr{\sqrt{2}M^D_{\nu}}{v_{\eta}},
		\hs  y_{N} =\frac{M_N}{\sqrt{2}v_{\phi}}.
		\ee
		Substituting into Eq.(\ref{yuk}), we get the corresponding results for the first and second terms on the right side as:

		\bea \label{lfvcoup1}
		g_{ab}\overline{\psi}_{aL}l_{bR}\rho&=&g_{ab}\left(\overline{\nu}_{aL}\rho^+_1+\overline{l}_{aL}\rho^0_2+\overline{(\nu_R^c)}_{aL}\rho^+_3\right)l_{bR}\crn
		&=& \frac{\sqrt{2}m_{l_a}}{v_\rho}\delta_{ab} \left(\overline{\nu}_{aL}\rho^+_1+\overline{l}_{aL}\rho^0_2+\overline{(\nu_R^c)}_{aL}\rho^+_3\right)l_{bR}\crn
		&\supset& \frac{\sqrt{2}m_{l_a}}{v_\rho}\delta_{ab} \left(\overline{\nu}_{aL}l_{bR}s_\alpha H^+_1+\overline{(\nu_R^c)}_{aL}l_{bR}c_{\theta_1}H^+_2 -\overline{l}_{aL}l_{bR}\left( h_2c_{\alpha_2}+h_1s_{\alpha_2}\right)\right), \crn
		\left( y_{\nu }^{D}\right) _{ab}\overline{\psi}_{aL}\eta N_{bR}&=&  \fr{\sqrt{2}\left(M^D_{\nu}\right)_{ab} }{v_{\eta}}\left( \overline{\nu}_{aL}\eta_1^0 + \overline{l}_{aL}\eta_2^- +\overline{\nu_R^c}_{aL}\eta_3^0 \right)  N_{bR}\crn
		&\supset&  \fr{\sqrt{2}\left( M^D_{\nu}\right)_{ab}}{v_{\eta}} \left( -\overline{\nu}_{aL}\left( h_1c_{\alpha_2}-h_2s_{\alpha_2} \right) + \overline{l}_{aL}H_1^-c_\alpha \right)  N_{bR}.
		\eea
		In Eq.(\ref{lfvcoup1}), we used supset ($\supset$) notation implying the omission of unnecessary interactions which appear when transfer from the initial interaction base to the physical base. The remainder are couplings related to LFV processes.
		
		In agreement with current experimental observations, charged leptons are assumed to be non-oscillating and active neutrinos generated masses by a type-I seesaw mechanism.
		We define transformations between the flavor basis  $\{l'_{aL,R}\}$,~$\left\lbrace n'_i\right\rbrace =\{\nu'_{aL},~(N'_{aR})^C\})$ and the mass basis $\{l_{aL,R}\}$,~$\left\lbrace n_i\right\rbrace =\{\nu_{aL},~(N_{aR})^C\}$:
		\bea
		l'_{aL}= l_{aL},  ~~l'_{aR}=l_{aR}, \;
		n'_{aL}=U^\nu_{ab}n_{bL},\;
		n^{'C}_{aL}=U^{\nu*}_{ab}n_{bL}^C\label{lepmixing}, \text{\,\,where } U^\nu \equiv \left(
		\begin{array}{cc}
			\mathcal{O}  & M^D_\nu \\
			(M^{D}_\nu)^T &  M_N   \\
		\end{array}
		\right), 
		\eea
		with $M^D_\nu$ and $ M_N$ have form as Eq.(\ref{ssmatrix}).\\
		From the above expansions, we show the lepton-flavor-violating couplings of this model in table. \ref{albga}.
		\begin{table}[h]
			\scalebox{0.82}{
				\begin{tabular}{|c|c|c|c|}
					\hline
					Vertex & Coupling & Vertex &Coupling \\
					\hline
					$\bar{\nu}_al_bH_1^{+},\,\bar{l}_b\nu_aH_1^{-}$&$i\sqrt{2}U^{\nu*}_{ba} \dfrac{m_{l_b}}{v_\rho}s_{\alpha}P_R,\,i\sqrt{2}U^{\nu}_{ba} \dfrac{m_{l_b}}{v_\rho}s_{\alpha}P_L$&$\bar{\nu}_al_bH_2^{+},~\nu_a\bar{l}_bH_2^{-}$&$i\sqrt{2}U^{\nu*}_{ba} \dfrac{m_{l_b}}{v_\rho}c_{\theta_1}P_R$, $i\sqrt{2}U^{\nu}_{ba} \dfrac{m_{l_b}}{v_\rho}c_{\theta_1}P_L$\\
					\hline	
					$\bar{N}_{a} l_bH_1^{+},~N_{a} \bar{l}_bH_1^{-}$ & $i\sqrt{2}\dfrac{(M_{\nu}^D)^*_{ab}}{v_\eta}c_{\alpha}P_L,~i\sqrt{2}\dfrac{(M_{\nu}^D)_{ab}}{v_\eta}c_{\alpha}P_R$ & $\bar{\nu}_a N_bh_2,~\bar{\nu}_a N_bh_1$ & $-i\sqrt{2}U^{\nu*}_{bc}\dfrac{(M_{\nu}^D)_{ab}}{v_\eta}s_{\alpha_2}P_L,~i\sqrt{2}U^{\nu*}_{bc}\dfrac{(M_{\nu}^D)_{ab}}{v_\eta}c_{\alpha_2}P_L$ \\
					\hline
					$\bar{l}_al_ah_1,~\bar{l}_al_ah_2$ & $\fr{i\sqrt{2}m_{l_a}}{v_\rho}s_{\al_2},~\fr{i\sqrt{2}m_{l_a}}{v_\rho}c_{\al_2}$ & $\bar{N}_aN_b\phi,~ H_\chi H_2^+Y^{\mu-}$ & $\frac{(M_N)_{ab}}{\sqrt{2}v_\phi c_{\al_\phi}},~\frac{igs_{\theta_1}}{2c_{\alpha_\phi}}\left( p_0-p_p\right) $\\
					\hline
					$\bar{\nu}_al_bW_\mu^{+},~\bar{\nu}_al_bY_\mu^{+}$&$\fr{ig}{\sqrt{2}}U^{\nu*}_{ba}\ga^\mu P_L$ & $\bar{l}_b\nu_aW_\mu^{-},~\bar{l}_b\nu_aYY_\mu^{-}$ & $\fr{ig}{\sqrt{2}}U^{\nu}_{ab}\ga^\mu P_L$\\
					\hline
					$W^{\mu+}W_\mu^{-}h_1,~W^{\mu+}W_\mu^{-}h_2$ & $\frac{igm_W}{2} \left(c_{\alpha } c_{\alpha _2}-s_{\alpha } s_{\alpha _2}\right), ~\frac{igm_W}{2}\left(c_{\alpha } s_{\alpha _2}-c_{\alpha _2} s_{\alpha }\right)$ & $Y^{\mu+}Y_\mu^{-}h_1,Y^{\mu+}Y_\mu^{-}h_2$ & $\frac{igm_W}{2} c_{\alpha } c_{\alpha _2} ,~\frac{igm_W}{2}c_{\alpha } s_{\alpha _2} $\\
					\hline
					$h_1H_1^{+}W^{\mu-},~h_2H_1^{+}W^{\mu-}$ & $-\frac{i g}{2}(p_0-p_p) \left(c_{\alpha _2} s_{\alpha }+c_{\alpha } s_{\alpha _2}\right),~ -\frac{i g}{2}(p_0-p_p) \left(c_{\alpha } c_{\alpha _2}+s_{\alpha } s_{\alpha _2}\right) $ &
					$h_1H_2^{+}Y^{\mu-},~h_2H_2^{+}Y^{\mu-}$ & $ -\frac{i g}{2}(p_0-p_p) c_{\alpha _2} c_{\theta _1} ,~-\frac{i g}{2}(p_0-p_p) c_{\theta _1} s_{\alpha _2}  $\\
					\hline
					$h_1 H_1^{+}H_1^{-},~h_1 H_2^{+}H_2^{-}$ & $-i \lambda_{h_1H_1H_1},~ -i \lambda_{h_1H_2H_2} $ & $h_2H_1^{+}H_1^{-},~h_2H_2^{+}H_2^{-}$ & $ -i \lambda_{h_2H_1H_1},~ -i \lambda_{h_2H_2H_2}$\\
					\hline
					
			\end{tabular}}
			\caption{Couplings relating with LFV process in 331ALP. All the couplings were only considered in the unitary gauge.} \label{albga}
		\end{table}
		
		where: 
		\bea \label{Higgs-sellcoup}
		\lambda_{h_2H^+_1H^-_1}&&= -v \left(-c_{\alpha _2}
		\left(\lambda _6+\lambda _9\right) c_{\alpha }^3+s_{\alpha } s_{\alpha
			_2} \left(2 \lambda _2+\lambda _9\right) c_{\alpha }^2-c_{\alpha _2}
		s_{\alpha }^2 \left(2 \lambda _2+\lambda _9\right) c_{\alpha }+s_{\alpha
		}^3 s_{\alpha _2} \left(\lambda _6+\lambda _9\right)\right) ,\crn
		\lambda_{h_2H^+_2H^-_2}&&=-s_{\alpha } s_{\alpha _2} v
		\left(\lambda _6 c_{\theta _1}^2+s_{\theta _1}^2 \lambda
		_5\right)+c_{\alpha } c_{\alpha _2} v \left(2 \lambda _2 c_{\theta
			_1}^2+s_{\theta _1}^2 \left(\lambda _5+\lambda
		_8\right)\right)+c_{\theta _1} s_{\theta _1} \left(c_{\alpha _2} v_{\chi
		} \lambda _8+s_{\alpha _2} v_{\phi } \lambda _{\phi }\right) ,\crn
		\lambda_{h_1H^+_1H^-_1}&&=v \left(s_{\alpha _2}
		\left(\lambda _6+\lambda _9\right) c_{\alpha }^3+c_{\alpha _2} s_{\alpha
		} \left(2 \lambda _2+\lambda _9\right) c_{\alpha }^2+s_{\alpha }^2
		s_{\alpha _2} \left(2 \lambda _2+\lambda _9\right) c_{\alpha }+c_{\alpha
			_2} s_{\alpha }^3 \left(\lambda _6+\lambda _9\right)\right) ,\crn
		\lambda_{h_1H^+_2H^-_2}&&=s_{\alpha _2}
		\left(c_{\theta _1} s_{\theta _1} v_{\chi } \lambda _8+c_{\alpha } v
		\left(2 \lambda _2 c_{\theta _1}^2+s_{\theta _1}^2 \left(\lambda
		_5+\lambda _8\right)\right)\right)+c_{\alpha _2} \left(s_{\alpha } v
		\left(\lambda _6 c_{\theta _1}^2+s_{\theta _1}^2 \lambda
		_5\right)-c_{\theta _1} s_{\theta _1} v_{\phi } \lambda _{\phi }\right).
		\eea
		In the framework of 331ALP, there are no trilinear couplings as $l_a\overline{l}_b\gamma$, ($a\neq b$). We can easily verify from the kinetic energy term: $\mathcal{L}^{kin}_{Lepton}=-i\bar{\psi}_{aL}D \!\!\!/_\mu  \psi_{aL}$. Therefore, decays of $(l_a \rightarrow l_b\gamma)$ appear only at the loop level. Based on table \ref{albga}, we can determine that the particles participating in the loop of these decays can be the following cases:$(\nu_a,W^\pm), (\nu_a,V^\pm), (\nu_a,H^\pm_{1,2}), \text{and}  (N_a,H_1^\pm) $. The left and right components of the amplitude of $(l_a \rightarrow l_b\gamma)$ are represented by the PV functions given in appendix \ref{appen_loop}.
		\bea
		\mathcal{D}_{(ab)L}&&=\mathcal{D}_{(ab)L}\left( m_{\nu_a},m_W\right)+\mathcal{D}_{(ab)L}\left( m_{\nu_a},m_V\right)+\mathcal{D}_{(ab)L}\left( m_{\nu_a},m_{H^\pm_{1,2}}\right)+\mathcal{D}_{(ab)L}\left( m_{N_a},m_{H^\pm_{1}}\right),\crn
		\mathcal{D}_{(ab)R}&&=\mathcal{D}_{(ab)R}\left( m_{\nu_a},m_W\right)+\mathcal{D}_{(ab)R}\left( m_{\nu_a},m_V\right)+\mathcal{D}_{(ab)R}\left( m_{\nu_a},m_{H^\pm_{1,2}}\right)+\mathcal{D}_{(ab)R}\left( m_{N_a},m_{H^\pm_{1}}\right).
		\eea
		The total branching ratios of the cLFV  processes can be shown based on Refs.~\cite{Crivellin:2018qmi,Thuc:2016qva, Hung:2021fzb,Quyet:2023dlo} and they have form as:
		\begin{equation}\label{eq_Gaebaga}
			\mathrm{Br}^{Total}(l_a\rightarrow l_b\gamma)\simeq \frac{48\pi^2}{ G_F^2} \left( \left|\mathcal{D}_{(ab)R}\right|^2 +\left|\mathcal{D}_{(ab)L}\right|^2\right) \mathrm{Br}(l_a\rightarrow l_b\overline{\nu_b}\nu_a),
		\end{equation}
		where $G_F=g^2/(4\sqrt{2}m_W^2)$, and for different charge lepton decays, we use experimental data $\mathrm{Br}(\mu\rightarrow e\overline{\nu_e}\nu_\mu)=100\%, \mathrm{Br}(\tau\rightarrow e\overline{\nu_e}\nu_\tau)=17.82\%, \mathrm{Br}(\tau\rightarrow \mu\overline{\nu_\mu}\nu_\tau)=17.39\% $ as given in Refs.\cite{Patrignani:2016xqp,Tanabashi:2018oca,Zyla:2020zbs}. \\
		
		The couplings in Tab.\ref{albga} are given based on the characteristics of the model under consideration. Therefore, we get the following interesting results: $\it{i}$) $\bar{N}_{a} l_bH_2^{+}=0$ leads to diagram 6) with only $H_1^\pm$ contributions, $\it{ii}$) $N_a$ does not interact with charged gauge bosons so diagram 8) does not give contributions for both $h_1$ and $h_2$ cases. Therefore, all the one-loop Feynman diagrams of $H\rightarrow l_al_b,\,\,(H\equiv h_1,\,h_2)$ decays are given in Fig.\ref{fig_hmt331} and the corresponding amplitudes are also shown in appendix \ref{appen_loop}.
		\begin{figure}[ht]
			\centering
			\begin{tabular}{cc}
				\includegraphics[width=14.0cm]{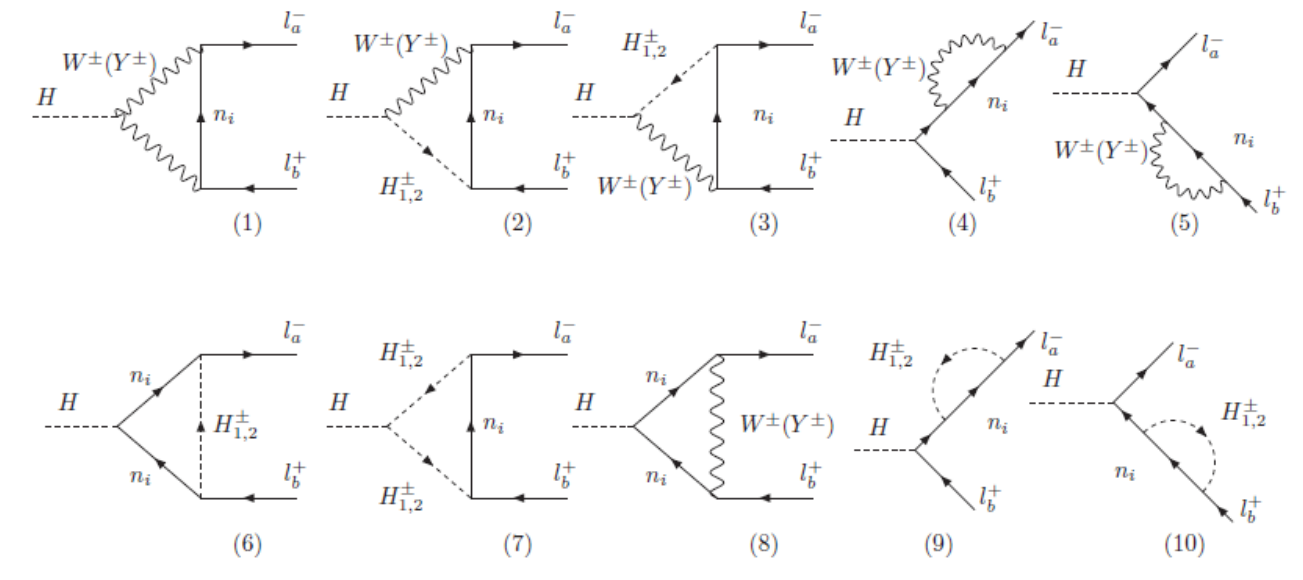} 
			\end{tabular}%
			\caption{ Feynman diagrams at one-loop order of $H \rightarrow l_a l_b$ decays in the unitary gauge, $H\equiv {h_1,h_2}$}
			\label{fig_hmt331}
		\end{figure}
		
		The partial width of $\mathrm{H}\rightarrow l_a^{\pm}l_b^{\mp}$ is
		\be
		\Gamma (\mathrm{H}\rightarrow l_al_b)\equiv\Gamma (\mathrm{H}\rightarrow l_a^{+} l_b^{-})+\Gamma (\mathrm{H} \rightarrow l_a^{-} l_b^{+})
		=  \fr{ m_{\mathrm{H}} }{8\pi }\left(\vert \Delta^{(ab)}_L\vert^2+\vert \Delta^{(ab)}_R\vert^2\right). \label{LFVwidth}
		\ee
		
		We use the conditions for external momentum as: $p^2_{a,b}=m^2_{a,b}$,\, $(p_a+p_b)^2=m^2_{\mathrm{H}}$ and $m^2_{\mathrm{H}}\gg m^2_{a,b}$,\, this leads to branching ratio of $\mathrm{H}\rightarrow l_a^{\pm}l_b^{\mp}$ decays can be given
		\bea  \mathrm{Br}(\mathrm{H} \rightarrow l_al_b)=\Gamma (\mathrm{H}\rightarrow l_al_b)/\Gamma^\mathrm{total}_{\mathrm{H}}, \label{brhmt} 
		\eea
		We need to know $\Gamma^\mathrm{total}_{\mathrm{H}}$, which can be addressed through current experimental data of the SM Higgs boson \cite{Hung:2019jue}, whose main contributions are,
		\bea \label{gam_tt1}
		\Gamma^\mathrm{total}_{\mathrm{H}}&=&\Gamma (\mathrm{H}\rightarrow b\overline{b})+\Gamma (\mathrm{H}\rightarrow c\overline{c})+ \Gamma (\mathrm{H}\rightarrow \tau \overline{\tau})+\Gamma (\mathrm{H}\rightarrow  W^*W)+\Gamma (\mathrm{H}\rightarrow Z^*Z)\crn
		&+&\Gamma (\mathrm{H}\rightarrow gg)+\Gamma (\mathrm{H}\rightarrow \gamma \gamma)+\Gamma (\mathrm{H}\rightarrow Z\gamma).
		\eea
	We are interested in the SM Higgs boson at $125.09~\gev$, the branching ratios of channels in Eq.(\ref{gam_tt1}) are given in table \ref{Br-H} \cite{Denner:2011mq} and the total decay width ($\Gamma^\mathrm{SM}_h$) is approximately $4.1\times 10^{-3}~\mathrm{GeV}$ \cite{CMS:2012hfp,CMS:2012zhx}. 
		
		\begin{table}[h] 
			\scalebox{1.25}{\begin{tabular}{|c|c|c|c|c|c|c|c|c|}
					\hline
					Channel & $\tau \overline{\tau}$ & $b \overline{b}$ & $c \overline{c}$ & $W^+W^-$ & $Z^*Z$ & $gg$ & $\gamma \gamma$ &$ Z\gamma$  \\ \hline
					Br($\%$) & $6.32$ & $57.7$ & $2.91$ & $21.5$ & $2.64$ & $8.57$ & $0.228$ &$ 0.154$  \\ \hline
					
			\end{tabular}}
			\caption{The branching ratios of SM Higgs boson decay with mass of $125.09~\gev$.}
			\label{Br-H}
		\end{table}
					
		For the CP-even Higgs bosons of the BSM, we can derive an estimate of $\Gamma^\mathrm{total}_{\mathrm{H}}$ by a similar method as shown in Ref.\cite{Barger:2012hv}. Accordingly, the partial decay width of each channel in Eq. (\ref{gam_tt1}) is given based on the branching ratios in table \ref{Br-H}  combined with a comparison to itself in SM. Applying this to the case $H\equiv h_1,~h_2$ in this model, we provide the couplings of $H$ with quarks related to Eq.(\ref{gam_tt1}) based on Eq.(\ref{yukintera}) as in table \ref{Coupling-Hf}. 
		
		\begin{table}[h] 
			\scalebox{1.25}{\begin{tabular}{|c|c|c|c|c|c|}
					\hline
					Couplings & $\overline{\tau}\tau $ & $ \overline{u_n}u_n$ & $\overline{u_3}u_3$ & $\overline{d_n}d_n$ & $\overline{d_3}d_3$  \\ \hline
					$h_1$ & $i\frac{m_\tau}{v_r}s_{\al_2}$ & $i\frac{m_{u_n}}{v_r}s_{\al_2}$ & $-i\frac{m_{u_3}}{v_e}c_{\al_2}$ & $-i\frac{m_{d_n}}{v_e}c_{\al_2}$ & $i\frac{m_{d_3}}{v_r}s_{\al_2}$  \\ \hline
					$h_2$ & $-i\frac{m_\tau}{v_r}c_{\al_2}$ & $-i\frac{m_{u_n}}{v_r}c_{\al_2}$ & $-i\frac{m_{u_3}}{v_e}s_{\al_2}$ & $-i\frac{m_{d_n}}{v_e}s_{\al_2}$ & $-i\frac{m_{d_3}}{v_r}c_{\al_2}$  \\ \hline
			\end{tabular}}
			\caption{The coupling of $H$ with fermions.}
			\label{Coupling-Hf}
		\end{table}
		
	Then, the partial decay widths of the first five channels in Eq.(\ref{gam_tt1}) are derived from a combination of  branching ratios in table \ref{Br-H} with the couplings in tables \ref{albga} and \ref{Coupling-Hf}. The last two channels in Eq.(\ref{gam_tt1}) are assumed to have the same branching ratios as SM. The total decay width is,
		\bea \label{gam_tt2}
		\Gamma^\mathrm{total}_{\mathrm{H}}&\simeq&\sum_f \mathrm{Br}(\mathrm{H} \rightarrow f\overline{f})\left( \frac{g_{Hff}}{g^{SM}_{hff}}\right)^2 \Gamma^{SM}_h+\sum_G \mathrm{Br}(\mathrm{H} \rightarrow G^*G)\left( \frac{g_{HGG}}{g^{SM}_{hGG}}\right)^2 \Gamma^{SM}_h\crn
		&+&\Gamma (\mathrm{H}\rightarrow gg)+\Gamma (\mathrm{H}\rightarrow \gamma \gamma)+\Gamma (\mathrm{H}\rightarrow Z\gamma),
		\eea
		where $G\equiv W^\pm,~Z$, $H\equiv h_1,~h_2$ and $f\equiv \tau, b, c$.
		
		The last three terms of Eq.(\ref{gam_tt2}) are loop-induced decays when considered in SM, but $h_{1,2}$ do not interact with exotic quarks so $\Gamma (\mathrm{H}\rightarrow gg)\simeq \Gamma (\mathrm{H}\rightarrow \overline{u_3}u_3),~u_3 \equiv t-quark$, the remaining two terms are given as,
		\bea \label{H-Zga}
		\Gamma (\mathrm{H}\rightarrow \gamma \gamma)=\mathrm{Br}(\mathrm{H}\rightarrow \gamma \gamma) \Gamma^{\mathrm{total}}_{\mathrm{H}}, ~\Gamma (\mathrm{H}\rightarrow Z \gamma)= \mathrm{Br}(\mathrm{H}\rightarrow Z \gamma) \Gamma^{\mathrm{total}}_{\mathrm{H}}.
		\eea
		By replacing Eq.(\ref{H-Zga}) with Eq.(\ref{gam_tt2}) and combining them with table \ref{Br-H} and table \ref{Coupling-Hf}, we can determine the total decay width of $h_1$ and $h_2$. Note that in Eq.(\ref{gam_tt2}), $\Gamma^\mathrm{SM}_{h}\simeq 4.1\times 10^{-3}~\mathrm{GeV}$ corresponding to $m_{h}=125.09\,[\mathrm{GeV}]$ \cite{CMS:2012hfp,CMS:2012zhx,Barger:2012hv} and the upper bound of the current experimental limit for $\mathrm{Br}(\mathrm{h} \rightarrow \mu \tau)$ is $10^{-3}$ as shown in Refs. \cite{Patrignani:2016xqp,Tanabashi:2018oca,Zyla:2020zbs}.
		
		\subsection{Numerical results}
		We use the well-known experimental parameters \cite{Zyla:2020zbs,Patrignani:2016xqp}: 
		the charged lepton masses $m_e=5\times 10^{-4}\,\mathrm{GeV}$,\,  $m_\mu=0.105\,\mathrm{GeV}$,\, $m_\tau=1.776\,\mathrm{GeV}$,\, the SM-like Higgs mass $m_{h}=125.1\,\mathrm{GeV}$,\,  the mass of the W boson $m_W=80.385\,\mathrm{GeV}$, the gauge coupling of the $SU(2)_L$ symmetry $g \simeq 0.651$ and $v=246\,\mathrm{GeV}$.
		The mixing angles act as both free and dependent parameters chosen below to investigate numerical of LFV process. But, a question is difficult as there are so many parameters. This problem can be solved by choosing $t_\alpha$ and $m_{H^\pm_2}$ as free parameters. Because, we can express $t_{\alpha_2}$ as dependent on $t_{\alpha_3}$, $t_{\alpha_\phi}$ and the $\mathrm{VeVs}$ based on the relationships in Eq.(\ref{ta2}).
		
		\bea
		t_{\alpha_2}&&=\frac{1}{4 c_{\alpha _3} v_{\eta } v_{\rho } \left(\lambda _6 v_{\eta
			} v_{\rho }+\lambda _{\phi } v_{\chi } v_{\phi }\right)}\left(4 \lambda _3 c_{\alpha _3}^2 v_{\eta } v_{\rho
		}^3-c_{\alpha _3}^2 \lambda _{\phi } v_{\eta }^2 v_{\chi } v_{\phi }-4 \lambda _3 v_{\eta }^3 v_{\rho }+\lambda _{\phi } v_{\rho }^2 v_{\chi } v_{\phi }\right. \crn
		&&\left. -\frac{1}{2} \sqrt{64 c_{\alpha _3}^2 v_{\eta }^2 v_{\rho }^2 \left(\lambda _6 v_{\eta } v_{\rho }+\lambda _{\phi } v_{\chi } v_{\phi }\right){}^2+\left(-8 \lambda _3 c_{\alpha _3}^2 v_{\eta } v_{\rho }^3+2
			c_{\alpha _3}^2 \lambda _{\phi } v_{\eta }^2 v_{\chi } v_{\phi }+8 \lambda _3 v_{\eta }^3 v_{\rho }-2 \lambda _{\phi } v_{\rho }^2 v_{\chi } v_{\phi }\right){}^2}\right),\crn
		t_{\alpha_3}&&=\frac{2 \lambda _1 v_{\chi }^4 c_{\alpha _{\phi }}+m_{A_5}^2 c_{\alpha } s_{\alpha } v_{\eta } v_{\rho } c_{\alpha _{\phi }}+ \sqrt{c_{\alpha _{\phi }}^2 \left(m_{A_5}^2 c_{\alpha } s_{\alpha } v_{\eta } v_{\rho }+2 \lambda _1 v_{\chi }^4\right){}^2+16 v_{\rho }^2 v_{\phi }^2 \left(m_{A_5}^2 c_{\alpha } s_{\alpha }
				v_{\eta }-\lambda _5 v_{\rho } v_{\chi }^2\right){}^2}}{4 v_{\rho } v_{\phi }
			\left(\lambda _5 v_{\rho } v_{\chi }^2-m_{A_5}^2 c_{\alpha } s_{\alpha } v_{\eta }\right)},\crn 
		t_{\alpha_\phi}&&=\frac{\sqrt{\lambda _{11}^2 v_{\chi }^2+\lambda _{10}^2 v_{\phi }^2}-\lambda _{10} v_{\phi }}{\lambda _{11} v_{\chi }}, \label{tmixangel}
		\eea
		where the VEVs expressed through other parameters such as: the mass of the new charged gauge boson ($m_V^\pm$), the mixing angle $\alpha$ ($t_\alpha=\frac{v_\eta}{v_\rho}$), in detail,  $v_{\chi }=\sqrt{\frac{4m_V^2}{g^2}-v_\rho^2},\, v_{\eta }=\frac{v t_a}{\sqrt{t_a^2+1}},\, v_{\rho }=\frac{v}{\sqrt{t_a^2+1}}$. The $\sin {\phi_i}$ and $\cos {\phi_i}$ functions of arbitrary angles $\phi_i$ can be related to $\tan {\phi_i}$ by the formulas: 	$s_{\phi_i}=\frac{t_{\phi_i}}{\sqrt{t_{\phi_i}^2+1}},\, c_{\phi_i}=\frac{1}{\sqrt{t_{\phi_i}^2+1}}$. Furthermore, the Yukawa and self-coupling of scalars constants are chosen to be smaller than $\sqrt{4\pi}$ and $4\pi$ respectively, to satisfy the limits of perturbation theory (Ref.\cite{Hung:2019jue}).\\ 
		Moreover, the interaction constants are chosen as dependent parameters based on Eqs.(\ref{massch},\ref{ta2}), the rest will be treated as free parameters and must be in the range of values that ensure the limits of perturbation theory ($\lambda_i \leq 4\pi$)   . These dependent parameters are listed as in Eq. (\ref{lambdas}).

		\bea \label{lambdas}
		\lambda_8&&=\frac{2 \left(m_{H_2}^2 v_{\chi }-v^2 m_{A_5}^2 c_{\alpha } s_{\alpha } v_{\eta }\right)}{v^2 v_{\rho } v_{\chi }^2}, \crn
		\lambda_9&&=\frac{2 \left(m_{H_1}^2-v^2 m_{A_5}^2 c_{\alpha } s_{\alpha }\right)}{v^2 v_{\eta } v_{\rho }},\crn
		\lambda _{\phi }&&=	-\frac{2 m_{A_5}^2 c_{\alpha } s_{\alpha }}{v_{\chi } v_{\phi }}, \crn
		\lambda_3&&=\frac{m_{A_5}^2 v_{\eta }^2 v_{\rho }^2-v^4 m_{A_5}^2+m_h^2 v^4+\sqrt{\left(m_{A_5}^2 \left(v_{\eta }^2 v_{\rho }^2-v^4\right)+m_h^2 v^4\right){}^2+3v^4 v_{\eta }^2 v_{\rho }^2 \left( m_h^2 m_{A_5}^2+\frac{3}{4} m_{A_5}^4- m_h^4\right)}}{3v^2 v_{\eta }^2 v_{\rho }^2}.
		\eea
		
		To numerically investigate the $l_a \rightarrow l_b\gamma$ and the LFVHDs, we next consider the parameterization of the mixing matrix $U^\nu$ of neutrinos, starting from the original relation of the seesaw mechanism. It has read:
		\bea 
		\hat{M}^\nu \equiv diag(m_{\nu_1},m_{\nu_2},m_{\nu_3})&&= U^\dagger_{PMNS}M^D_\nu M_N^{-1}(M_\nu^D)^TU_{PMNS}, \\ \label{para-Unu}
		U_{PMNS}\hat{M}^\nu U^\dagger_{PMNS}&& = M^D_\nu M_N^{-1}(M_\nu^D)^T. \label{para-Unu1}
		\eea
		
		By using the neutrino oscillation data as shown in Ref.\cite{Zyla:2020zbs,Patrignani:2016xqp, Tanabashi:2018oca, ParticleDataGroup:2024cfk, JUNO:2021vlw}: $s^2_{12}=0.307,\; s^2_{23}=0.558,\; s^2_{13}=0.0219,\crn
		\Delta m^2_{21}= 7.53\times 10^{-5}\;\mathrm{ eV^2},\hs  \Delta m^2_{32}= 2.455\times 10^{-3}\; \mathrm{eV^2}$, and choosing the Dirac and Majorana phases of the $U_{PMNS}$ in their simplest form (Refs.\cite{Quyet:2023dlo, Hung:2022vqx, Hue:2021zyw}), we can get the right part of Eq.(\ref{para-Unu1}) in the known form. Furthermore, $M_\nu^D$ must be antisymmetric so it contains only six unknowns. Combined with $M_N$ in diagonal form, we get three more unknowns. There are a total of nine unknowns on the right part, so Eq.(\ref{para-Unu1}) can be solved. That means we have successfully parametrized $U^\nu$.\\
		
\begin{figure}[ht]
			\centering
			\begin{tabular}{c}
				\includegraphics[width=7.0cm]{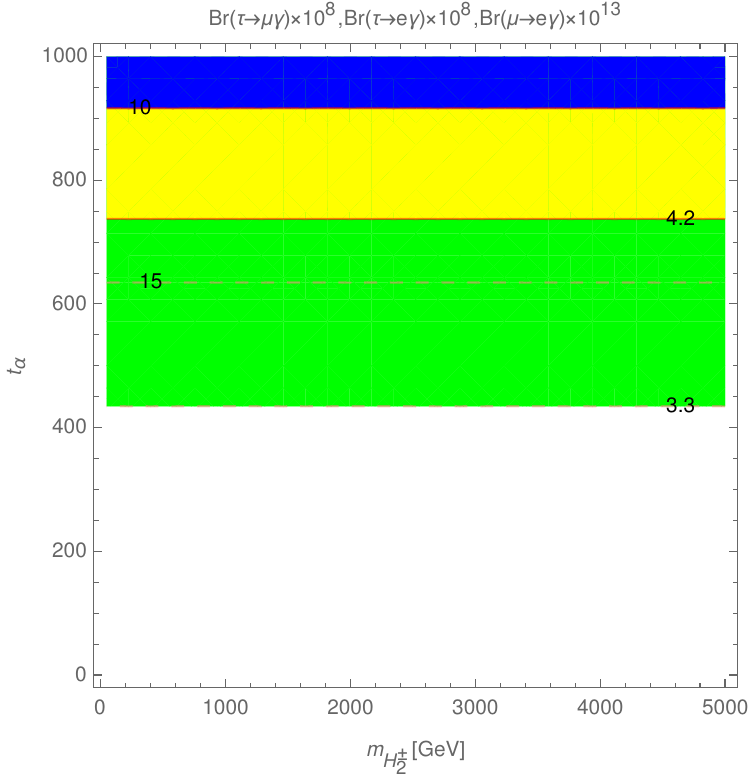} 
			\end{tabular}%
			\caption{ Contour plots of $l_a \rightarrow l_b\gamma$ decays in plane of ($t_\alpha, m_{H^\pm_2}$).}
			\label{fig_lalb}
		\end{figure}

		We investigate the numerical results of $l_a \rightarrow l_b\gamma$ in the parameter space $10^{-3}\leq t_\alpha \leq 10^3$ and $300\leq m^\pm_{H_2} \leq 5 \times 10^3\, \mathrm{GeV}$. We obtain that $\tau \rightarrow e\gamma$ is always smaller than the experimental limit ($4.4 \times 10^{-8}$). The part of the parameter space exceeding the upper limit of $\tau \rightarrow \mu\gamma$ is indicated in green, with pink dashed lines corresponding to values $3.3\times 10^{-8}$, $15\times 10^{-8}$ of $\mathrm{Br}(\tau \rightarrow \mu\gamma)$. Meanwhile, the yellow and blue markers are used to represent the parameter space in which $\mathrm{Br}(\mu \rightarrow e\gamma)$ take values larger than the upper bound. We can find the curves for the values $ 4.2\times 10^{-13}$ and $ 10\times 10^{-13}$ of $\mathrm{Br}(\mu \rightarrow e\gamma)$ in there. Therefore, the space region allowed to satisfy the experimental constraints on $l_a \rightarrow l_b\gamma$ decays is the uncolored part in Fig.\ref{fig_lalb}.

\begin{figure}[ht]
			\centering
			\begin{tabular}{cc}
				\includegraphics[width=6.5cm]{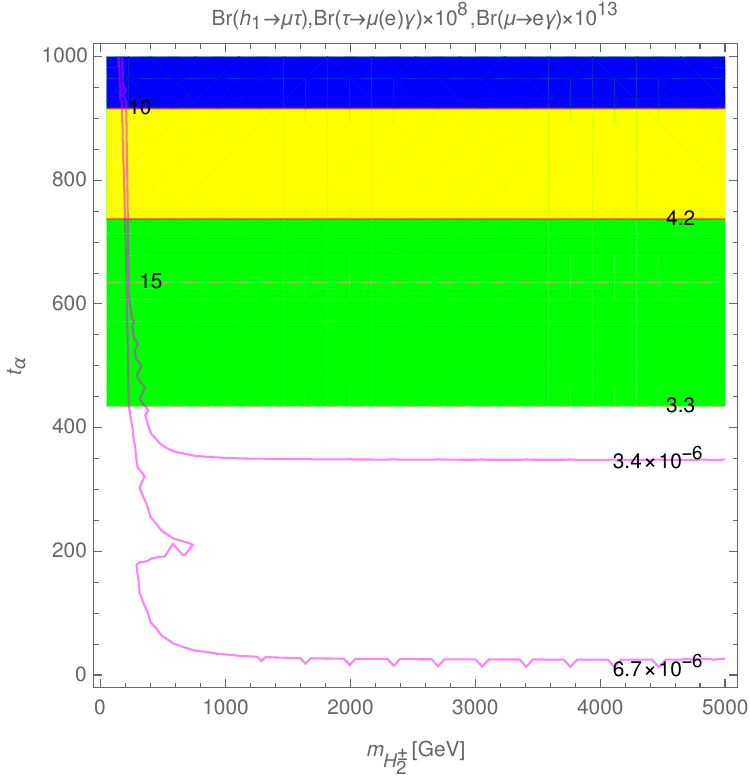} &\includegraphics[width=6.5cm]{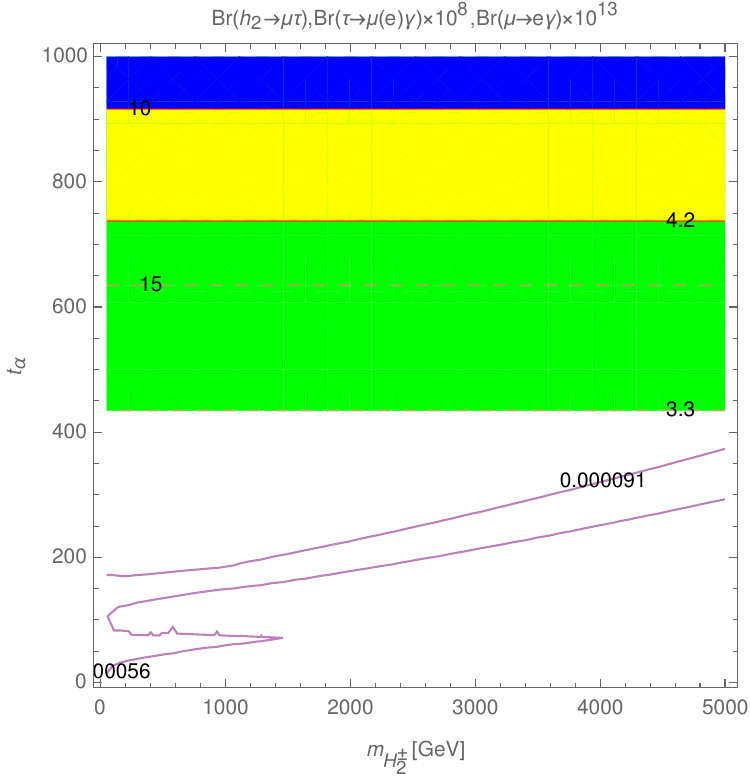}
			\end{tabular}%
			\caption{ Plots of $\mathrm{Br}(h_1 \rightarrow \mu \tau)$  (left panel) and $\mathrm{Br}(h_2 \rightarrow \mu \tau)$  (right panel) in plane of ($t_\alpha, m_{H^\pm_2}$).}
			\label{fig_density}
		\end{figure}

		Based on the analytical form of the amplitudes given in appendix \ref{appen_loop}, we investigate the values of   $\mathrm{Br}(h_1 \rightarrow \mu \tau)$ and   $\mathrm{Br}(h_2 \rightarrow \mu \tau)$, which are given in Fig.\ref{fig_density}. We can see that, most of the values obtained by $\mathrm{Br}(h_1 \rightarrow \mu \tau)$, in left panel, and $\mathrm{Br}(h_1 \rightarrow \mu \tau)$, in right panel, are within the current experimental limits ($\leq 10^{-3}$). 
		
		To illustrate, we represent the signals of $\mathrm{Br}(h_1 \rightarrow \mu \tau)$ and   $\mathrm{Br}(h_2 \rightarrow \mu \tau)$, in the allowed space region in Fig.\ref{fig_density}. On the left panel of Fig.\ref{fig_density}, $\mathrm{Br}(h_1 \rightarrow \mu \tau)$ may be get values of $3.4 \times 10^{-6}$ and $6.7 \times 10^{-6}$ while  $\mathrm{Br}(h_2 \rightarrow \mu \tau)$ may be get values of $9.1 \times 10^{-5}$ and $5.6 \times 10^{-5}$ in right panel. These values all satisfy the present experimental limits.\\

		Next, we predict the mass of the new CP-even Higgs boson based on guluon-gluon fusion at the LHC and data analysis of CMS. The signal of the Higgs boson product proton-proton collision is modeled assuming a narrow width of the Higgs boson \cite{LHCHiggsCrossSectionWorkingGroup:2013rie}. In this case, the partons can be gluon or quarks \cite{Langenfeld:2012ti,Das:2025tuk}. Thus, the hadronic cross section depends on the parton luminosity and partonic cross section as follow: \cite{Langenfeld:2012ti, Das:2025tuk},
		\bea \label{pp-collision}
		\sigma \left(pp \rightarrow H \right) = \sum_{p_i,p_j=\left\lbrace g,~q\right\rbrace }\mathcal{L}_{p_ip_j}\left( m_H^2 \right) \hat{\sigma}\left(p_ip_j \rightarrow H \right),  
		\eea
		where $g$ and $q$ are denoted to gluon and quark, respectively. 
	
	    As shown in Eq.(\ref{pp-collision}), Higgs boson production in proton-proton collisions at the LHC involves partons, which include gluons and quarks. However, due to their rather large mass, the density of quarks within the proton is significantly lower than that of gluons. Consequently, the gluon parton distribution plays a main role in Higgs production, with the gluons eventually leading a quark loop. Therefore, we approximate the hadronic cross section when partons are gluons, its form as \cite{Anastasiou:2016cez, Davies:2019wmk},
	    \bea
		\sigma \left(pp \rightarrow H \right)\simeq \sigma \left(gg \rightarrow H \right) = \mathcal{L}_{gg}\left( m_H^2 \right) \hat{\sigma}\left(gg \rightarrow H \right).  
		\eea
		The parton luminosity relate to parton distribution function as:
		\bea
		\mathcal{L}_{gg }\left( m_H^2 \right) = \int_{\frac{m_H^2}{s}}^1\frac{dx_1}{x_1}\int_{x_1}^1\frac{dx_2}{x_2} \left[ F_{g}\left(\frac{x_1}{x_2}, m_H^2  \right) F_{g}\left(x_2, m_H^2  \right)\right] \delta \left(x_1x_2- \frac{m_H^2}{s}\right),  
		\eea
		where $F_{g}\left(x, m_H^2  \right)$ is known as the parton distribution functions of gluon and $\sqrt{s}$ to be the collider center-of mass energy, $\hat{\sigma}$ is the partonic cross section expanded in terms of the strong interaction constant ($\al_s$) as follows Refs.\cite{Langenfeld:2012ti, Das:2025tuk}:
		\bea \label{hatcr}
		\hat{\sigma}\left(gg \rightarrow H \right) = \al^2_s \left[ \hat{\sigma}^{(0)} \left(gg \rightarrow H  \right)+  \al_s \hat{\sigma}^{(1)} \left(gg \rightarrow H  \right)+... \right]. 
		\eea
		The loop corrections involving intermediate quarks are the primary source of Higgs boson production via the gluon-gluon fusion. However, the t-quark makes the largest contribution to this process, as Higgs boson couples about 35 times more strongly to the t- quark than to the next-heaviest fermion, the b- quark . Consequently, other contributions can be neglected, and only the t-quark contribution need be considered. So, the first order approximation in Eq.(\ref{hatcr}) is then expressed as follows \cite{Langenfeld:2012ti, Harlander:2010su}:
		\bea
		\hat{\sigma}^{(0)} \left(gg \rightarrow H  \right) = \frac{9G_Fm^2_H \zeta^2}{1152\sqrt{2}\pi}\left| 1+(1-\zeta)f(\zeta)\right| ^2 , 
		\eea
    	where $G_F$ is Fermi constant, $\zeta=\frac{4m^2_t}{m^2_H}$ and
    		\bea
    		f\left(\zeta \right) = \left\lbrace 
    		\begin{array}{c}
    			\arcsin^2{\frac{1}{\sqrt{\zeta}}},\hs \zeta \geq 1 \\
    			\frac{1}{2}\left[ln\frac{1-\sqrt{1-\zeta}}{1+\sqrt{1-\zeta}}-i\pi \right]^2, \hs \zeta < 1 \\
    		\end{array}.
    		\right.  
    		\eea
		In dilepton decay, if the decay width is assumed to be very narrow then the cross section of these processes can be given approximately as:
		\bea
		\sigma \left(gg \rightarrow H \rightarrow \overline{l}_il_j \right) \simeq \sigma \left(gg \rightarrow H  \right) \times \mathrm{Br} \left(H \rightarrow \overline{l}_il_j \right). 
		\eea
		By analyzing the data with center of mass energy at 13 $\mathrm{TeV}$ and integrated luminosity of $35.9~\mathrm{fb}^{-1}$ of detector, CMS has shown that $\sigma \left(gg \rightarrow H  \right) \times \mathrm{Br} \left(H \rightarrow \mu\tau \right)$ can take values from $51.9~ (57.4)~\mathrm{fb}$ to $1.6 ~(2.1)~\mathrm{fb}$  at $95 \%$ confidence level \cite{CMS:2019pex}. To optimize the processing of experimental data, CMS divided into two scales corresponding to low mass 200-450 $\mathrm{GeV}$ and high mass 450-900 $\mathrm{GeV}$.
		\begin{figure}[ht]
			\centering
			\begin{tabular}{c}
				\includegraphics[width=10.0cm]{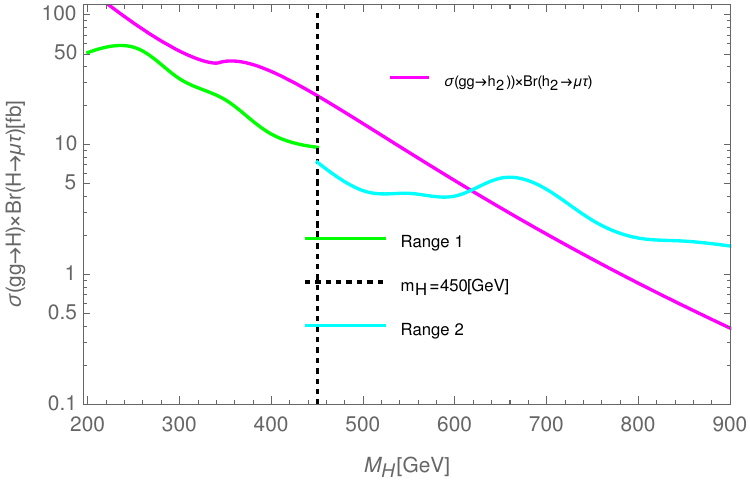}
			\end{tabular}%
			\caption{ Plots of $\sigma \left(gg \rightarrow h_2 \right) \times \mathrm{Br} \left(h_2 \rightarrow \mu\tau \right)$ (magenta line) depend on $M_H$, the green and cyan lines depict the CMS's observations for the combined decay modes of $H \rightarrow \mu\tau$ (based on Ref.~\cite{CMS:2019pex}) for the low and high mass ranges, respectively.}
			\label{fig_com3}
		\end{figure}
		
		As results, we describe the combined two processes $\sigma \left(gg \rightarrow H  \right) \times \mathrm{Br} \left(H \rightarrow \mu\tau_h \right)$ and  $\sigma \left(gg \rightarrow H  \right) \times \mathrm{Br} \left(H \rightarrow \mu\tau_e \right)$ at $95\%$ CL as the green line and cyan line in Fig.~\ref{fig_com3}, the dotted black represents the boundary between the two mass scales.
		
		To predict the masses of the new CP-even Higgs bosons, $h_2$, the parton distribution functions are chosen to be the non-relativistic form (Breit-Wigner distribution functions), the partonic cross section is approximated at the lowest level, and for simplicity $\hat{\sigma}\left(gg \rightarrow H \right)\sim \frac{G_Fm^2_H \zeta^2}{\sqrt{2}\pi}\left| 1+(1-\zeta)f(\zeta)\right| ^2 $. So, we investigate signal of $\sigma \left(gg \rightarrow h^0_2 \right) \times \mathrm{Br} \left(h^0_2 \rightarrow \mu\tau \right)$ as magenta line in Fig. \ref{fig_com3}, and  mass of $h_2$ is predicted to be greater $630 \gev$.
		
		\section{$Z^{\prime}$ boson}
		\label{heavy} 
		In 331ALP, the heavier neutral gauge boson, $Z'$, has the mass and physical state given by Eq.(\ref{mngauge}) in section \ref{model}. The interaction of $Z'$ with fermions originates from the kinetic term of spinor fields with the specific form as follows:
		\bea
		\mathcal{L}^{Z^\prime}_{NC} \supset \overline{f}\gamma^\mu\left(g_V^{Z^\prime}(f)+g_A^{Z^\prime}(f) \gamma^5\right)f Z^\prime_\mu,
		\eea
		where $g_V^{Z^\prime}$ and $g_A^{Z^\prime}$ are given in table \ref{G-av}.
		\begin{table}[h] 
			\scalebox{0.92}{\begin{tabular}{|c|c|c|c|c|c|c|c|c|}
					\hline
					f & $\nu_e, \nu_\mu, \nu_\tau$ & $e, \mu, \tau$ & $u,c$ & $d,s$ & $t$ & $b$ & $U_3$ &$ D_{n} $  \\ \hline
					$g^{Z^\prime}_V$ & $g\left( \frac{s_{\beta_1}}{4\sqrt{3}}-\frac{2tc_{\beta_1}}{3\sqrt{6}}\right) $ & $g\left( \frac{s_{\beta_1}}{4\sqrt{3}}-\frac{2tc_{\beta_1}}{3\sqrt{6}}\right) $ & $g\left( \frac{s_{\beta_1}}{4\sqrt{3}}+\frac{tc_{\beta_1}}{3\sqrt{6}}\right) $ & $g\left( \frac{s_{\beta_1}}{4\sqrt{3}}-\frac{tc_{\beta_1}}{6\sqrt{6}}\right) $ & $g\left( \frac{s_{\beta_1}}{4\sqrt{3}}+\frac{tc_{\beta_1}}{2\sqrt{6}}\right) $ & $\frac{gs_{\beta_1}}{4\sqrt{3}}$ & $g\left( -\frac{s_{\beta_1}}{2\sqrt{3}}+\frac{tc_{\beta_1}}{3\sqrt{6}}\right) $ &$ -\frac{gs_{\beta_1}}{2\sqrt{3}}$  \\ \hline
					$g^{Z^\prime}_A$ & $g\left( \frac{s_{\beta_1}}{4\sqrt{3}}-\frac{tc_{\beta_1}}{3\sqrt{6}}\right) $ & $g\left( \frac{s_{\beta_1}}{4\sqrt{3}}-\frac{tc_{\beta_1}}{3\sqrt{6}}\right) $ & $g\left( \frac{s_{\beta_1}}{4\sqrt{3}}+\frac{tc_{\beta_1}}{3\sqrt{6}}\right) $ & $g\left( \frac{s_{\beta_1}}{4\sqrt{3}}-\frac{tc_{\beta_1}}{6\sqrt{6}}\right) $ & $g\left( \frac{s_{\beta_1}}{4\sqrt{3}}+\frac{tc_{\beta_1}}{6\sqrt{6}}\right) $ & $g\left( \frac{s_{\beta_1}}{4\sqrt{3}}-\frac{tc_{\beta_1}}{3\sqrt{6}}\right) $ & $g\left( -\frac{s_{\beta_1}}{2\sqrt{3}}+\frac{tc_{\beta_1}}{3\sqrt{6}}\right) $ &$g\left( -\frac{s_{\beta_1}}{2\sqrt{3}}-\frac{tc_{\beta_1}}{3\sqrt{6}}\right) $  \\ \hline
			\end{tabular}}
			\caption{The couplings of $Z^{\prime}$ with fermions, with $\beta_1$ angle defined in Eq.(\ref{betangle}) and $t=\frac{g_X}{g}=\frac{3\sqrt{2}t_W}{\sqrt{3-t^2_W}}$ as shown in Ref.\cite{Ninh:2005su}.}
			\label{G-av}
		\end{table}
		
		We also use the kinetic energy of the scalar fields, such as Eq.(\ref{j231}), to derive the couplings non-zero of $Z^\prime$ with gauge and Higgs bosons as table \ref{Zp-boson}.
		\begin{table}[h] 
			\scalebox{1.25}{\begin{tabular}{|c|c|}
					\hline
					Vertex  & Coupling \\ \hline
					$Z^\prime H^+_1H^-_1$  & $-\frac{2g \left(c_{\alpha }^2 \left(\sqrt{4 c_W^2-1} c_{\beta _1} \left(\sqrt{3}-3 s_{\beta _1}\right)-6 s_W s_{\beta _1} s_{\beta _2}\right)-s_{\alpha
						}^2 \left(\sqrt{4 c_{W}^2-1} c_{\beta _1} \left(3 s_{\beta _1}+\sqrt{3}\right)+12 s_W s_{\beta _1} s_{\beta _2}\right)\right)}{6 \sqrt{4 c_{W}^2-1}} $ \\ \hline
					$Z^\prime H^+_2H^-_2$  & $-\frac{2g  \left(2 \left(\sqrt{12 c_{W}^2-3} c_{\beta _1}-6 s_{W} s_{\beta _1} s_{\beta _2}\right) c_{\theta _1}^2+\left(\sqrt{4 c_{W}^2-1} c_{\beta _1}
						\left(\sqrt{3}-3 s_{\beta _1}\right)-6 s_{W} s_{\beta _1} s_{\beta _2}\right) s_{\theta _1}^2\right)}{6 \sqrt{4 c_{W}^2-1}} $ \\ \hline
					$Z^\prime H^+_1W^-$  & $\frac{1}{3} gm_W c_{\alpha } s_{\alpha } \left(\sqrt{3} c_{\beta _1}+\frac{3 s_{W} s_{\beta _1} s_{\beta _2}}{\sqrt{4 c_{W}^2-1}}\right) $ \\ \hline
					$Z^\prime H^+_2Y^-$  & $-\frac{g^2 \left(c_{\alpha } c_{\theta _1} \left(\sqrt{4 c_{W}^2-1} c_{\beta _1} \left(3 s_{\beta _1}+\sqrt{3}\right)-24 s_{W} s_{\beta _1} s_{\beta _2}\right)
						v+\left(\sqrt{4 c_{W}^2-1} c_{\beta _1} \left(3 s_{\beta _1}+\sqrt{3}\right)+12 s_{W} s_{\beta _1} s_{\beta _2}\right) s_{\theta _1} v_{\chi }\right)}{12 \sqrt{4 c_{W}^2-1}}$ \\ \hline
			\end{tabular}}
			\caption{The couplings of $Z^{\prime}$ with charged and gauge bosons.}
			\label{Zp-boson}
		\end{table}
		
		The experimental LHC, with the final product of the dilepton at $\sqrt{s}=13~\mathrm{TeV}$, suggests that the process in which $Z^{\prime}$ is mediated has a cross section as the product of the hadronic cross section and the branching ratio \cite{CMS:2021ctt}. 
		\begin{align} \label{g11}
			\sigma \left(pp \rightarrow Z^\prime \rightarrow \overline{l}l \right) \simeq \sum_q\int_\tau^1\frac{dx_1}{x_1}\int_\tau^1\frac{dx_2}{x_2} \xi_q\left(\frac{x_1}{x_2}, m_Z^2  \right) \overline{\xi}_q\left(x_2, m_Z^2  \right)\hat{\sigma}\left(\overline{q}qi \rightarrow Z \right)\times \mathrm{Br} \left(Z \rightarrow \overline{l}l \right),  	
		\end{align}
		where $\xi_q$ and $\overline{\xi}_q$ are the parton distribution functions of quarks and antiquarks, respectively. The partonic cross section and the branching ratio are given as:
		\bea \label{g12}
		\hat{\sigma}\left(\overline{q}q \rightarrow Z^\prime \right) &=& \frac{\pi g^2}{3 c^2_W}\sum_i \left[  \left(g^{Z^\prime}_V(q_i) \right)^2+  \left(g^{Z^\prime}_A(q_i) \right)^2 \right],\crn
		\mathrm{Br} \left(Z^\prime  \rightarrow \overline{l}l \right)&=&\frac{\Gamma \left( Z^\prime \rightarrow \overline{l}l \right)}{\Gamma ^{tot}_{ Z^\prime}}, 	
		\eea
		with the decay channel products being dileptons, the decay width has the form as in Eq.(\ref{dilepton} ), while the total decay width has to take into account the gauge and Higgs boson contributions, denoted as $G$ and $H$, as listed in Eq.(\ref{proGH}).
		\begin{align} \label{dilepton}
			\Gamma \left( Z^\prime \rightarrow \overline{l}l \right) \simeq \frac{m_{Z^\prime}}{12\pi} \frac{\pi g^2}{c^2_W} \sum_{e,\mu,\tau,\nu,N_a} \sqrt{1-\frac{4m^2_l}{m^2_{Z^\prime}}} \left[  \left(g^{Z^\prime}_V(l) \right)^2\left( 1+\frac{2m^2_l}{m^2_{Z^\prime}}\right)  +  \left(g^{Z^\prime}_A(l) \right)^2\left( 1-\frac{4m^2_l}{m^2_{Z^\prime}}\right)  \right], 
		\end{align}
		
		\begin{align} \label{proGH}
			\Gamma ^{tot}_{ Z^\prime}\simeq \Gamma \left( Z^\prime \rightarrow \overline{f}f \right)+\Gamma \left( Z^\prime \rightarrow H^*H\right)+\Gamma \left( Z^\prime \rightarrow GH\right)+\Gamma \left( Z^\prime \rightarrow G^*G\right). 
		\end{align}
		Because $Z$ and $Z^\prime$ do not mix as shown in Eq.(\ref{mngauge}), $\Gamma (Z^\prime \rightarrow W^+W^-)$ is canceled out according to Ref.\cite{Perez:2004jc}. Furthermore, $m_X > m_{Z^\prime}/2$ means $\Gamma (Z^\prime \rightarrow X^*X,~X \equiv V^\pm, X^{0*}, X^0$ also do not contribute to the total width of $Z^\prime$. The remaining components are in the form as follows \cite{Ninh:2005su,cl,Perez:2004jc}:
		\begin{align} \label{p-fusion}
			\Gamma \left( Z^\prime \rightarrow \overline{f}f \right) \simeq \frac{m_{Z^\prime}}{12\pi} \frac{\pi g^2}{c^2_W} \sum_{f} N_C(f)\sqrt{1-\frac{4m^2_f}{m^2_{Z^\prime}}} \left[  \left(g^{Z^\prime}_V (f)\right)^2\left( 1+\frac{2m^2_f}{m^2_{Z^\prime}}\right)  +  \left(g^{Z^\prime}_A (f)\right)^2\left( 1-\frac{4m^2_f}{m^2_{Z^\prime}}\right)  \right].  	
		\end{align}
		\begin{align} \label{p-fusn}
			\Gamma \left( Z^\prime \rightarrow H^*H\right)  \simeq \frac{m_{Z^\prime}}{48\pi}\sum_{H}g_{Z'HH}^2 \left(1-\frac{4m^2_H}{m^2_{Z^\prime}} \right)^{3/2}.  	
		\end{align}
		\begin{align} \label{p-fsn}
			\Gamma \left( Z^\prime \rightarrow GH\right)  \simeq \sum_{G,H} \frac{g_{Z'GH}^2}{24\pi}\frac{E^2-m_G^2}{m_{Z^\prime}^2} \left(2+\frac{E^2}{m^2_{G}} \right), 	
		\end{align}
		where $E=\frac{m^2_{Z^\prime}+m^2_G-m^2_H}{2m_{Z^\prime}}$ is denoted as energy of the gauge boson at the final state and the factors are taken from tables \ref{G-av} and \ref{Zp-boson}.	
		
\begin{figure}[ht]
			\centering
			\begin{tabular}{c}
				\includegraphics[width=10cm]{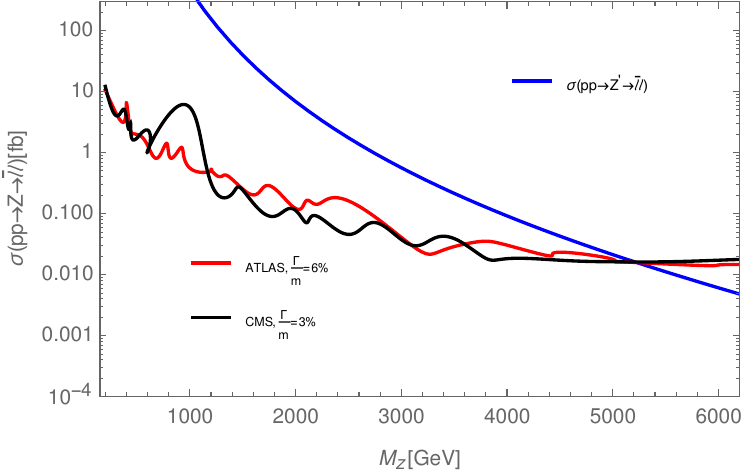} 
			\end{tabular}%
			\caption{ The dependence of $\sigma \left(pp \rightarrow Z^\prime \rightarrow \overline{l}l \right)$ on $M_Z$ (blue line). The red line represent the upper limits of cross section by ATLAS with $\frac{\Gamma}{m}=6\%$ \cite{ATLAS:2019erb}, The black line represent the upper limits of cross section by CMS with $\frac{\Gamma}{m}=3\%$ at $95\%$CL \cite{CMS:2021ctt}  }
			\label{fig_contour}
		\end{figure}

		Based on the analytical form in Eqs.(\ref{g11},~\ref{g12}) and the known data, as given in Section \ref{scalarfields}, we show the dependence of $\sigma \left(pp \rightarrow Z^\prime \rightarrow \overline{l}l \right)$ on the mass of $m_Z$ as blue line in Fig.\ref{fig_contour}. The red line is used to depict the upper limits of dilepton production cross section when ATLAS gives the result at an integrated
		luminosity of $139~ \mathrm{ fb^{-1}}$ \cite{ATLAS:2019erb}, while the black line is based on the upper bound given by CMS for $95\%$CL and $\sqrt{s}=13~\mathrm{TeV}$ \cite{CMS:2021ctt}. Therefore, to satisfy the constraints, we can get $m_{Z^\prime}\geqslant 5.1~\mathrm{TeV}$, of course, below this mass value, $Z^\prime$ is excluded. The limit of $Z^\prime$ mass in this work is very suitable for non-supersymmetry models, similar to $Z^{\prime}_{\psi}$ in grand unified theories based on $E_6$ gauge group or $Z^{\prime}_{SSM}$ in generalized sequential model \cite{ATLAS:2019erb,CMS:2021ctt}. However, a major difference occurs with supersymmetry models, where the limit of $m_{Z^{\prime}}$, as mentioned above, is much lower \cite{Athron:2015tsa,Cvetic:1997mbh}. 
		\section{Dark matter}
		\label{darkmatter}
		From the symmetry breaking scheme in (\ref{scheme}), we see that $Z_2$ is the residual symmetry of the model, so {\bf odd}-$Z_2$ particles can be considered as dark matter candidates. These include: $\eta,\,\rho,\,u_{aR},\,d_{nR},\,l_{aR},\,N_{aR}$ according to (\ref{oct1}). But $\eta$ and $\rho$ directly break electroweak scale, and $u_{aR},\,d_{nR},\,l_{aR}$ are components that create ordinary matter, so only $N_{aR}$ remains as the best choice to consider as dark matter candidates in the model.
		
\begin{figure}[ht]
			\centering
			\begin{tabular}{c}
				\includegraphics[width=6.5cm]{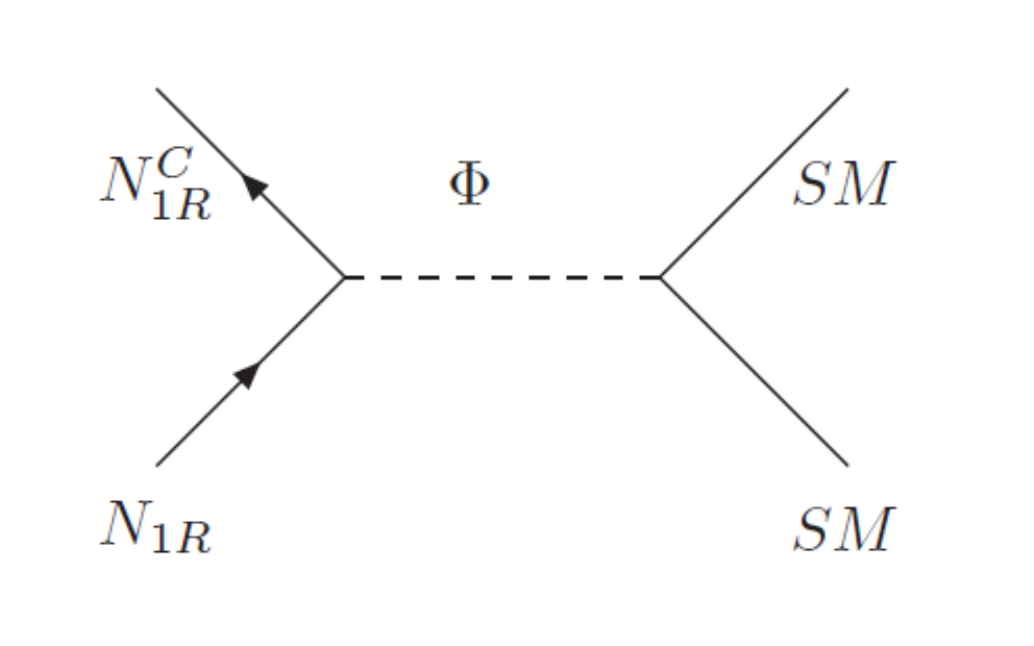} 
			\end{tabular}%
			\caption{ Dominant contributions to annihilation of dark matter.}
			\label{figgDM}
		\end{figure}

		Among $N_{aR}$, we assume $N_{1R}$ is the lightest. So, $N_{1R}$ is considered dark matter and its pair annihilates in to the SM particles in the early universe. Because $N_{1R}$ only direct interacts with $\Phi$ according to Eq.(\ref{yukintera}), and $\Phi$ can couple with SM particles as Eq.(\ref{poten3}), the contribution to the annihilation of dark matter particles mainly comes from the s-channel exchange diagrams mediated by $\Phi$. The diagram is shown in Figure \ref{figgDM}.	
		Although $\Phi$ mixes with $R_\chi$ and $R_\phi$ so it can interact with fermions arcooding to Eq.(\ref{PhysCPeven}), the mixing angles are very small, as Eq.(\ref{ta2}). Therefore, the SM product of the dark matter pair annihilation channel does not include fermions but only gauge bosons. It means $N_{1R}^CN_{1R} \rightarrow h_1h_1,~ ZZ$. Hence, product of the annihilation cross-section of the dark matter pair with relative velocity is approximately \cite{Cannoni:2015wba,CMS:2021ctt},
		\bea
		\langle \sigma v_{rel}\rangle _{N_{1R}^CN_{1R} \rightarrow all}&&=2\sigma_{\Lambda}F_{-4}(x_{f})\simeq \frac{3}{x_{f}} \sigma_{\Lambda} \crn
		&&\simeq \frac{3m^2_{N_{1R}}}{4\pi x_{f}}\left\lbrace  \frac{\left(g_{\Phi N_{1R}^CN_{1R}} \right)^2}{4m^2_{N_{1R}}-m^2_{\Phi} } \times \left[\left(g_{\Phi ZZ} \right)^2 +\left(g_{\Phi h_1h_1} \right)^2\right] \right\rbrace , \label{crdm}
		\eea
		where $x_f\simeq 20.25$ \cite{Bertone:2004pz} and $g_{\Phi N_{1R}^CN_{1R}}=\left(y_N \right)_{11} $, while on the contrary $g_{\Phi ZZ}=0$, $g_{\Phi h_1h_1}=\frac{v_\phi}{2\sqrt{2}}\left(\lambda_{12} s_{\al_2}^2+\lambda_{13} c_{\al_2}^2 \right) $ are determined from Eq.(\ref{j231}) and Eq.(\ref{poten3}), respectively. 
		
		We employed the relic abundance of dark matter as in Ref.\cite{Planck:2018vyg} follows:
		\bea
		\Omega_{DM} h^2\simeq \frac{0.1~\mathrm{pb}}{\langle \sigma v_{rel}\rangle _{N_{1R}^CN_{1R} \rightarrow all}}.
		\eea
		\begin{figure}[ht]
			\centering
			\begin{tabular}{c}
				\includegraphics[width=8cm]{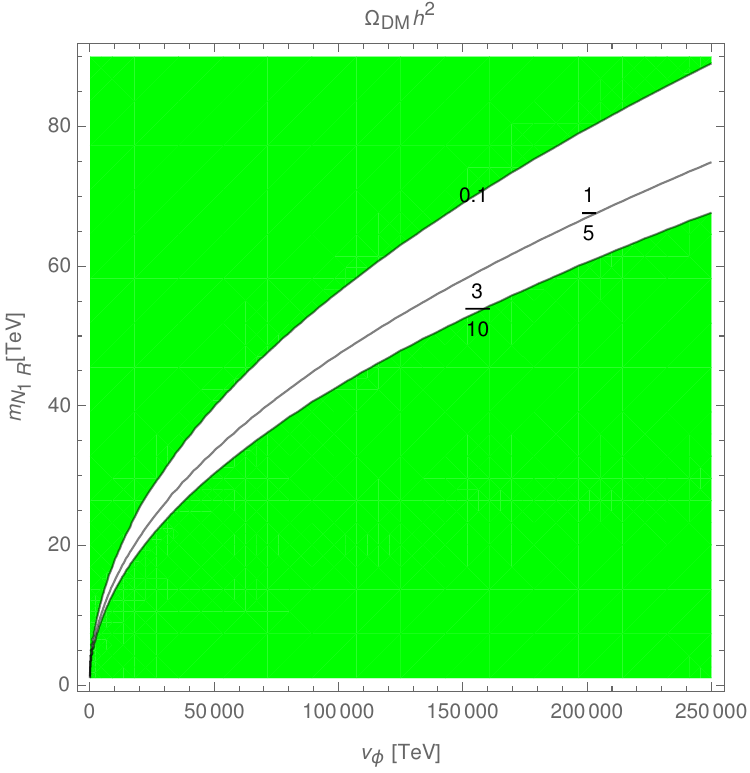} 
			\end{tabular}%
			\caption{ Contour of the dark matter relic density on plane of ($v_\Phi,~m_{N_{1R}}$).}
			\label{figgDMp}
		\end{figure}
		We represent the dark matter relic density, $0.1\leq\Omega_{DM} h^2 \leq 0.3$ \cite{Bahcall:1999xn}, in the plane of ($v_\Phi,~m_{N_{1R}}$) as shown in Figure \ref{figgDMp}. We recall that for the PQ symmetry to not be broken under the influence of quantum gravity effects and to ensure the appearance of a naturally light, stable axion then $v_\phi\simeq 10^{11}~\gev$ ( as mentioned in Section 2). Therefore, $v_\phi$ is considered the breaking scale of the axion in the model. As a result in Figure \ref{figgDMp}, the colorless is the parameter space region that satisfies current observations, in which there is a relationship between the mass of dark matter and the breaking scale of axion.
		
		\section{Conclusions}
		\label{Conclusion}	
		In the 3-3-1 model with axionlike particles, we show the region of the parameter space where both cLFV and LFVHDs are satisfied experimental limits. These are allowed parameter spaces where $\mathrm{Br}(h_1 \rightarrow \mu\tau)$ has a value about of $10^{-6}$ and $\mathrm{Br}(h_2 \rightarrow \mu\tau)$ reaches a value approximately of $10^{-5}$. Based on gluon-gluon fusion at the LHC, we investigate the signals of cross sections in the allowed parameter space, and provide predictions of  $m_{h_2}\geq 630 ~\mathrm{GeV}$. Based on the search for high-mass dilepton resonances at ATLAS and CMS, we take the numerical result of $\sigma \left(pp \rightarrow Z^\prime \rightarrow \overline{l}l \right)$ and show that the mass of $Z^\prime$-boson is $m_{Z^\prime}\geq 5.1 ~\mathrm{TeV}$. The hypothesis of the existence of residual symmetry $Z_2$ after spontaneous symmetry breaking stages allows that only $N_{aR}$ are dark matter candidates among the  odd -$Z_2$ particles. Investigating the relic density of dark matter within experimentally permissible limits, we also established a relationship between the mass of dark matter and the breaking scale of axion.

		\section*{Acknowledgments}
		H.T. Hung would like to thank A.V. Bednyakov for the helpful comments to improving this article. 
		
		\appendix
		\section{Master integrals.}
		\label{appen_PV}
		In this section, we use the Passarino-Veltman (PV) functions as mentioned in Ref.~\cite{Passarino:1978jh} to establish the contributions at one-loop order of the Feynman diagrams in Fig.~\ref{fig_hmt331}. By introducing the formulas $D_0=k^2-M_0^2+i\delta$, $D_1=(k-p_1)^2-M_{1}^2+i\delta$ and $D_2=(k+p_2)^2-M_2^2+i\delta$, with $\delta$ is  infinitesimally a  positive real quantity, we give:
		\bea
		A_{0}(M_n)
		&\equiv &\frac{\left(2\pi\mu\right)^{4-D}}{i\pi^2}\int \frac{d^D k}{D_n}, \hs B^{(12)}_0 \equiv \frac{\left(2\pi\mu\right)^{4-D}}{i\pi^2}\int \frac{d^D k}{D_1D_2},\hs
		B^{(1)}_0 \equiv \frac{\left(2\pi\mu\right)^{4-D}}{i\pi^2}\int \frac{d^D k}{D_0D_1}, \crn
		B^{(2)}_0 &\equiv& \frac{\left(2\pi\mu\right)^{4-D}}{i\pi^2}\int \frac{d^D k}{D_0D_2}, 
		\hs C_0 \equiv  C_{0}(M_0,M_1,M_2) =\frac{1}{i\pi^2}\int \frac{d^4 k}{D_0D_1D_2},
		\label{scalrInte}\eea
		where $n=1,2$, $D=4-2\epsilon \leq 4$ is the dimension of the integral, while $~M_0,~M_1,~M_2$ indicate the masses of virtual particles in the loops. We also suppose  $p^2_1=m^2_{1},~p^2_2=m^2_{2}$ for external particle. The tensor integrals are
		\bea
		A^{\mu}(p_n;M_n)
		&=&-\frac{i\left(2\pi\mu\right)^{4-D}}{\pi^2}\int \frac{k^{\mu}\times d^D k }{D_n}=A_0(M_n)p_n^{\mu},\crn
		B^{\mu}(p_n;M_0,M_n)&=& -\frac{i\left(2\pi\mu\right)^{4-D}}{i\pi^2}\int \frac{k^{\mu}\times d^D k}{D_0D_n}\equiv B^{(n)}_1p^{\mu}_n,\crn
		B^{\mu}(p_1,p_2;M_1,M_2)&=& -\frac{i\left(2\pi\mu\right)^{4-D}}{i\pi^2}\int \frac{k^{\mu}\times
			d^D k}{D_1D_2}\equiv B^{(12)}_1p^{\mu}_1+B^{(12)}_2p^{\mu}_2,\crn
		C^{\mu}(M_0,M_1,M_2)&=&-\frac{i}{\pi^2}\int \frac{ k^{\mu}\times d^4 k}{D_0D_1D_2}\equiv  C_1 p_1^{\mu}+C_2 p_2^{\mu},\crn
		C^{\mu \nu}(M_0,M_1,M_2)&=&-\frac{i}{\pi^2}\int \frac{k^{\mu}k^{\nu} \times d^4 k }{D_0D_1D_2}\equiv  C_{00}g^{\mu \nu}+C_{11} p_1^{\mu}p_1^{\nu}+C_{12} p_1^{\mu}p_2^{\nu}+C_{21} p_2^{\mu}p_1^{\nu}+C_{22} p_2^{\mu}p_2^{\nu},\crn
		\label{oneloopin1}\eea
		where $A_0$, $B^{(n)}_{0,1}$, $B^{(12)}_{n}$ and $C_{0,n}, C_{mn}$   are PV functions.  It is noted that only $A_0$, $B^{(n)}_{0,1}$, $B^{(12)}_{n}$ contain the divergence which is denoted as
		\be \Delta_{\epsilon}\equiv \frac{1}{\epsilon}+\ln4\pi-\gamma_E, \label{divt}\ee with $\gamma_E$ is the  Euler constant.  
		
		Using the technique as mentioned in Ref.~\cite{Hue:2017lak}, the divergent parts are shown 
		\bea  \mathrm{Div}[A_0(M_n)]&=& M_n^2 \Delta_{\epsilon}, \hs  \mathrm{Div}[B^{(n)}_0]= \mathrm{Div}[B^{(12)}_0]= \Delta_{\epsilon}, \crn
		\mathrm{Div}[B^{(1)}_1]&=&\mathrm{Div}[B^{(12)}_1] = \frac{1}{2}\Delta_{\epsilon},  \hs  \mathrm{Div}[B^{(2)}_1] = \mathrm{Div}[B^{(12)}_2]= -\frac{1}{2} \Delta_{\epsilon}.  \label{divs2}\eea
		
		Therefore, the above PV functions can be written in form:
		\be  A_0(M)= M^2\Delta_{\epsilon}+a_0(M),\,\, B^{(n)}_{0,1}= \mathrm{Div}[B^{(n)}_{0,1}]+ b^{(n)}_{0,1}, \,\,  B^{(12)}_{0,1,2}= \mathrm{Div}[B^{(12)}_{0,1,2}]+ b^{(12)}_{0,1,2}, \label{B01i}\ee
		where $a_0(M), \,\,b^{(n)}_{0,1}, \,\, b^{(12)}_{0,1,2} $ are finite parts and have a specific form defined as Ref.~\cite{Thuc:2016qva,Hung:2022kmv, Phan:2016ouz} for LFVHDs.
		\section{Analytic formulas at one-loop order to $l_a \rightarrow l_ab\gamma$ and $\mathrm{H} \rightarrow l_il_j$.}
		\label{appen_loop}
		In this section, we will give analytical forms at one-loop order for decays of $l_a \rightarrow l_b\gamma$ and $\mathrm{H} \rightarrow l_il_j$. Based on the couplings mentions in Tab.\ref{albga}, decays of $(l_a \rightarrow l_b\gamma)$ appear only at the loop level. These contributions include: $\mathcal{D}^{CH}_{L,R}(\nu_a,W^\pm), \mathcal{D}^{CH}_{L,R}(\nu_a,V^\pm), \mathcal{D}^{CH}_{L,R}(\nu_a,H^\pm_{1,2}), \text{and}\,  \mathcal{D}^{CH}_{L,R}(N_a,H_1^\pm) $. They can read \cite{Hung:2021fzb,Hung:2022kmv}:
		
		\begin{align}
			\mathcal{D}^{CH}_{L,R}(\nu_a,W^\pm)&=-\frac{eg^2}{32\pi^2m_W^2} \sum_{n=1}^9U^{\nu*}_{an}U^{\nu}_{bn}F(t_{nW}),\crn
			\mathcal{D}^{CH}_{L,R}(\nu_a,Y^\pm)&=  -\frac{eg^2}{32\pi^2m_Y^2} \sum_{n=1}^9U^{\nu*}_{(a+3)n}U^{\nu}_{(b+3)n}F(t_{nY}), \crn
			\mathcal{D}^{CH}_{L,R}(\nu_a,H^\pm_{1,2})&=-\frac{eg^2f_s}{16\pi^2m_W^2} \sum_{k=1}^9\left[\frac{ \lambda^{L,s*}_{ak}  \lambda^{L,s}_{bk}}{m^2_{H^{\pm}_s}}\times\frac{1-6t_{ks} +3 t^2_{ks} +2t^3_{ks} -6t^2_{ks} \ln(t_{ks})}{12 (t_{ks}-1)^4} \right.\crn
			&+\left. \frac{m_{n_k} \lambda^{L,s*}_{ak} \lambda'^{R,s}_{bk}}{m^2_{H^{\pm}_s}}\times \frac{-1 +t_{ks}^2 -2t_{ks} \ln(t_{ks})}{2(t_{ks}-1)^3} \right],\crn
			\mathcal{D}^{CH}_{L,R}(N_a,H_1^\pm)	&=-\frac{eg^2f_s}{16\pi^2m_W^2} \sum_{k=1}^9\left[\frac{ \lambda^{L,s*}_{ak}  \lambda^{L,s}_{bk}}{m^2_{H^{\pm}_s}}\times\frac{1-6t_{ks} +3 t^2_{ks} +2t^3_{ks} -6t^2_{ks} \ln(t_{ks})}{12 (t_{ks}-1)^4} \right.\
			\label{DRlalbga}
		\end{align}
		Where, we use symbols
		\begin{align}
			t_{ks}=\left(\frac{m_{n_k}}{m_{H^\pm_s}} \right) ^2, ~	t_{nW}=\left(\frac{m_{n_n}}{m_{W^\pm}} \right) ^2, ~	t_{nY}=\left(\frac{m_{n_n}}{m_{Y^\pm}} \right) ^2.
		\end{align}
		
		The one-loop factors of the diagrams in Fig.(\ref{fig_hmt331}) are given below. We used the same calculation techniques as shown in \cite{Thuc:2016qva} and the symmetry coefficients of the diagrams are quoted as in Refs.\cite{Hue:2010xr, Dong:2009db}. We give $m_{l_i} \equiv  m_1$ and $ m_{l_j} \equiv m_2$.
		{\small\bea  \mathcal{M}^{(1)}_L(m_F,m_Y)
			&=& m_1\left\{ \frac{1}{2m_Y^3}\left[m_F^2(B^{(1)}_1-B^{(1)}_0-B^{(2)}_0)\right.\right.\crn
			\hs &-&\left.\left. m_2^2B^{(2)}_1 + \left(2m_V^2+m^2_{H}\right)m_F^2\left(C_0-C_1\right)\right]\right.\crn
			&&\left.-\left(2m_Y+\frac{m_1^2-m_2^2}{m_Y}\right) C_1 +
			\left(\frac{m_1^2-m^2_{H}}{m_Y}+ \frac{m_2^2 m^2_{H}}{2m_Y^3}\right)C_2\right\}, \label{EfvvL} \\
			\mathcal{M}^{(1)}_R(m_F,m_Y)&=& m_2\left\{\frac{1}{2 m_Y^3}\left[-m_F^2\left(B^{(2)}_1+ B^{(1)}_0 + B^{(2)}_0 \right) \right.\right.\crn
			&+& \left.\left.  m_1 ^2 B^{(1)}_1  +   (2m_V^2+m^2_{H}) m_F^2(C_0+C_2)\right] \right.\crn
			&&\left.+\left(2m_Y-\frac{m_1^2-m_2^2}{m_Y}\right)C_2-\left( \frac{m_2^2-m^2_{H}}{m_Y}+ \frac{m_1^2 m^2_{H}}{m_Y^3}\right)C_1\right\},  \label{EfvvR}
			\eea
			\bea
			&& \mathcal{M}^{(2)}_L(a_1,a_2,v_1,v_2,m_F,m_Y,m_{H^\pm})\crn&=&
			m_1\left\{-\fr{a_2}{v_2} \fr{m_F^2}{m_Y^2}\left(B^{(1)}_1-B^{(1)}_0\right)   + \fr{a_1}{v_1}m_2^2\left[2 C_1-\left(1+ \fr{m^2_{H^\pm}-m^2_{H}}{m_Y^2}\right) C_2\right]\right.\crn
			&&\left.+\fr{a_2}{v_2}\frac{m_F^2}{m_Y^2}\left[(m^2_{H^\pm}-m^2_{H}+m_Y^2)C_0+(m^2_{H}-m^2_{H^\pm}+m_Y^2)C_1\right]\right\}, \label{EfvhL} \\
			&& \mathcal{M}^{(2)}_R(a_1,a_2,v_1,v_2,m_F,m_Y,m_{H^\pm})\crn &=&m_2\left\{\fr{a_1}{v_1}\left[\fr{m_1^2B^{(1)}_1-m_F^2B^{(1)}_0}{m_Y^2} +\left(\frac{}{}m_F^2C_0-m_1^2C_1+2 m_2^2C_2\right.\right.\right.\crn
			&&\left.\left.+2(m^2_{H}-m_2^2)C_1-  \fr{m^2_{H^\pm}-m^2_{H}}{m_Y^2}\left(m^2_FC_0-m_1^2C_1\right)\right)\right]\crn &&+\left.+\fr{a_2}{v_2}\frac{m_F^2}{m_Y^2}\left[-2m_Y^2C_0-(m^2_{H}-m^2_{H^\pm}+m_Y^2)C_2\right] \right\},   \label{EfvhR}
			\eea
			\bea
			&& \mathcal{M}^{(3)}_L(a_1,a_2,v_1,v_2,m_F,m_{H^\pm},m_Y)\crn&=& m_1\left\{\fr{a_1}{v_1}\left[-\fr{m_2^2B^{(2)}_1+m_F^2B^{(2)}_0}{m_Y^2} +\left(\frac{}{}m_F^2C_0-2m_1^2C_1+ m_2^2C_2\right.\right.\right.\crn
			&&\left.\left.+2(m_1^2-m^2_{H})C_2+  \fr{m^2_{H}-m^2_{H^\pm}}{m_Y^2}\left(m^2_FC_0+m_2^2C_2\right)\right)\right] \crn &&\left.+\fr{a_2}{v_2} m_F^2\left(-2C_0+\fr{m_Y^2+m^2_{H^\pm}-m^2_{H}}{m_Y^2}C_1 \right)\right\}, \label{EfhvL} \\
			&& \mathcal{M}^{(3)}_R(a_1,a_2,v_1,v_2,m_F,m_{H^\pm},m_Y)\crn&=& m_2 \left\{\fr{a_2}{v_2} \fr{m_F^2}{m_Y^2}\left(B^{(2)}_1+B^{(2)}_0\right)
			+ \fr{a_1}{v_1}m_1^2\left[-2 C_2+\left(1+ \fr{m^2_{H^\pm}-m^2_{H}}{m_Y^2}\right) C_1\right]\right.\crn
			&&\left.+\fr{a_2}{v_2}\frac{m_F^2}{m_Y^2}\left[(m_Y^2+m^2_{H^\pm}-m^2_{H})C_0+(m^2_{H^\pm}-m^2_{H}-m_Y^2)C_2 \right]\right\}.   \label{EfhvR}
			\eea
			\bea
			\mathcal{M}^{(4+5)}_L(m_F,m_Y)&=& \fr{-m_1m_2^2}{m_Y^3(m_1^2-m_2^2)}\left[\left(2m_Y^2+m_F^2\right) \left(B^{(1)}_1 +B^{(2)}_1 \right) \right. \crn&+&\left.m_1^2 B^{(1)}_1 +m_2^2 B^{(2)}- 2m_F^2\left(B^{(1)}_0-B^{(2)}_0\right)\right],  \label{DfvL} \\
			\mathcal{M}^{(4+5)}_R(m_F,m_Y)&=& \frac{m_1}{m_2}\mathcal{M}^{(4+5)}_L(m_F,m_Y), \label{DfvR}\eea
			\bea
			\mathcal{M}^{(6)}_L(a_1,a_2,v_1,v_2,m_F,m_{H^\pm})&=&\frac{ m_1m^2_F }{v_2}\times \left[\fr{a_1^2}{v_1^2}m_2^2(2C_2+C_0)+\fr{a_2^2}{v_2^2}m_F^2(C_0-2C_1) \right.\crn
			&+& \left.\fr{a_1a_2}{v_1v_2} \left(B^{(12)}_{0}+(m_F^2+m^2_{H^\pm}+m_2^2)C_0
			+\frac{}{}2m_2^2C_2-(m_1^2+m_2^2)C_1\right)\right],\crn  \label{EhffL} \\
			\mathcal{M}^{(6)}_R(a_1,a_2,v_1,v_2,m_F,m_{H^\pm})&=& \frac{m_2 m^2_F}{v_2}\times\left[\dfrac{a_1^2}{v_1^2}m_1^2(C_0-2C_1)+\fr{a_2^2}{v_2^2}m_F^2(C_0+2C_2)\right.\crn
			&+&\left. \fr{a_1a_2}{v_1v_2}\left(B^{(12)}_{0}+(m_F^2+m^2_{H^\pm}+m_1^2)C_0 -2m_1^2C_1+(m_1^2+m_2^2)C_2\right)\right], \crn\label{EhffR} \eea
			\bea
			\mathcal{M}^{(7)}_L(a_1,a_2,v_1,v_2,m_F,m_{H^\pm})&=&  m_1v_2\left[ \fr{a_1a_2}{v_1v_2}m_F^2C_0+\fr{a^2_2}{v^2_2}m_F^2C_1-\fr{a^2_1}{v^2_1}m_2^2C_2\right] , \crn \label{EfhhL} \\
			\mathcal{M}^{(7)}_R(a_1,a_2,v_1,v_2,m_F,m_{H^\pm})&=& m_2v_2\left[ \fr{a_1a_2}{v_1v_2}m_F^2C_0+\fr{a^2_1}{v^2_1}m_1^2C_1-\fr{a^2_2}{v^2_2}m_F^2C_2 \right],\crn \label{EfhhR} \eea
			\bea
			\mathcal{M}^{(8)}_L(m_Y,m_F)&=&\frac{m_1m^2_F}{m_Y^3}\times\left[\left(B^{(12)}_{0}
			+B^{(1)}_1 -(m_1^2+m_2^2-2m_F^2)C_1\right)-\left( C_0-4C_1\right)m_Y^2 \right],\crn  \label{EvffL} \\
			\mathcal{M}^{(8)}_R(m_Y,m_F)&=&
			\frac{m_2 m^2_F}{m_Y^3} \times\left[ \left(B^{(12)}_{0} -B^{(2)}_1 +(m_1^2+m_2^2-2m_F^2)C_2\right)-\left( C_0+4C_2\right)m_Y^2  \right],\crn \label{EvffR}\eea
			
			\bea
			\mathcal{M}^{(9+10)}_L(a_1,a_2,v_1,v_2,m_F,m_{H^\pm})&=& \fr{m_1}{(m_1^2-m_2^2)v_1}\left[m^2_2 \left(m^2_1\fr{a_1^2}{v_1^2}+m^2_F\fr{a^2_2}{v^2_2} \right)
			\left(B_1^{(1)}+B_1^{(2)}\right) \right.\crn
			&&\left.  \hspace{1.8 cm}+m^2_F\fr{a_1a_2}{v_1v_2}\left(2m^2_2B_0^{(1)}-(m^2_1 +m^2_2)B_0^{(2)}\right) \right], \label{DfhL} \\
			\mathcal{M}^{(9+10)}_R(a_1,a_2,v_1,v_2,m_F,m_{H^\pm})&=&  \fr{m_2}{(m_1^2-m_2^2)v_1}\left[ m^2_1 \left(m^2_2\fr{a_1^2}{v_1^2}+m^2_F\fr{a^2_2}{v^2_2} \right)\left(B_1^{(1)}+B_1^{(2)}\right)\right.\crn
			&&\left.  \hspace{1.8 cm}+m^2_F\fr{a_1a_2}{v_1v_2}\left(-2m^2_1B_0^{(2)}+(m^2_1 +m^2_2)B_0^{(1)}\right)\right]. \label{DfhR} \eea}
		In the 7th diagram of Fig.(\ref{fig_hmt331}) we always get that $\mathcal{M}^{(7)}_{L,R}(a_1,a_2,v_1,v_2,m_F,m_H)$ are finite because $C_k$ ($k=0,1,2$) do not contain divergent functions.


\begin{thebibliography}{99}
			
			
		\bibitem{Calcuttawala:2018wgo} Calcuttawala, Zaineb and Kundu, Anirban and Nandi, Soumitra and Kumar Patra, Sunando, Phys. Rev. D \textbf{97} (9): 095009 (2018).
			
			\bibitem{Yan:2025luk} Yan, Xin-Shuai and Hu, Zhen-Qing and Chang, Qin and Yang, Ya-Dong, Phys. Rev. D \textbf{112} (5): 055026 (2025).	
		 
	      \bibitem{Lychkovskiy:2010ue} Lychkovskiy, Oleg and Blinnikov, Sergei and Vysotsky, Mikhail, Eur. Phys. J. C \textbf{67} (7): 213--227 (2010).
     
			
			\bibitem{Patrignani:2016xqp} Patrignani, C. and others, Chin. Phys. \textbf{C40} (10): 100001 (2016).	
			
			\bibitem{Tanabashi:2018oca}
			M.~Tanabashi \textit{et al.}, Phys. Rev. D \textbf{98} (3): 030001 (2018). 
			
			\bibitem{ParticleDataGroup:2024cfk} Navas, S. and others, Phys. Rev. D \textbf{110} (3): 030001 (2024). 
			
			\bibitem{ATLAS:2019erb} Aad, Georges and others, Phys. Lett. B \textbf{796}:68-87 (2019). 
			
			\bibitem{Zyla:2020zbs}
			P.~A.~Zyla \textit{et al.} [Particle Data Group],
			PTEP \textbf{2020} (8): 083C01 (2020). 
			
			
			\bibitem{BaBar:2009hkt} Aubert, Bernard and others, Phys. Rev. Lett., \textbf{104}: 021802 (2010). 
			
			
			\bibitem{MEG:2016leq} Baldini, A. M. and others, Eur. Phys. J. C, \textbf{76}: 434 (2016). 	
			
			\bibitem{CMS:2018ipm} Sirunyan, Albert M and others, JHEP.\textbf{06}: 120 (2018).
			
			\bibitem{ATLAS:2019pmk} Aad, Georges and others,  Phys. Lett. B\textbf{800}: 135069 (2020). 
			
			\bibitem{ATLAS:2019old} Aad, Georges and others, Phys. Lett. B.\textbf{801}: 135148 (2020). 
			
			\bibitem{Mizukoshi:2010ky}
			Mizukoshi, J.K. and de S.Pires, C.A. and Queiroz, F.S. and Rodrigues da Silva, P.S.,
			Phys.\ Rev.\ D {\bf 83}:065024  (2011). 
					
			\bibitem{Dias:2005yh}
			Dias, Alex G. and de S.Pires, C.A. and Rodrigues da Silva, P.S.,
			Phys. Lett. B {\bf 628}: 85-92 (2005). 
						
			\bibitem{Hung:2022vqx} Hung, H. T. and Arbuzov, A. B., Chin.Jour. Phys \textbf{102}: 264-284 (2026).
			
			\bibitem{Hung:2022kmv} Hung, H. T. and Binh, D. T. and Quyet, H. V., Chin. Phys. C.\textbf{46} (12): 123104 (2022). 
			
			\bibitem{PhysRevD.47.2918} J. C. Montero, F. Pisano, and V. Pleitez, Phys. Rev. \textbf{D47} (7): 2918 (1993). 
			
			\bibitem{Boucenna:2015zwa} Boucenna, Sofiane M. and Valle, Jose W. F., Phys. Rev. \textbf{D92}: 053001 (2015). 
			
			\bibitem{Hernandez:2013hea} 
			Carcamo Hernandez, Antonio Enrique and Martinez, R. and Ochoa, F., EPJC {\bf C76}: 634 (2016). 
			
			\bibitem{Nguyen:2018rlb}Nguyen, T.Phong and Le, T. Thuy and Hong, T.T. and Hue, L.T., Phys. Rev. \textbf{D97}: 073003 (2018).
			
			\bibitem{Hung:2021fzb} Hung, H. T. and Tham, N. T. and Hieu, T. T. and Hang, N. T. T., PTEP.\textbf{2021} (8): 083B01 (2021). 	
			
			\bibitem{Ferreira:2019qpf} Ferreira, Manoel M. and de Melo, Tessio B. and Kovalenko, Sergey and Pinheiro, Paulo R. D. and Queiroz, Farinaldo S., Eur. Phys. J. C \textbf{79}, (11): 955 (2019). 
			
			\bibitem{CarcamoHernandez:2019pmy}
			Carcamo Hernandez, A.E. and Marchant Gonzalez, Juan and Saldana-Salazar, U.J.,Phys. Rev. {\bf D100}: 035024 (2019). 
			
			\bibitem{Catano:2012kw} Catano, M.E. and Martinez, R and Ochoa, F., Phys. Rev. \textbf{D86}: 073015 (2012). 
			
			\bibitem{Hernandez:2014lpa}
			Carcamo Hernandez, A. E. and Catano Mur, E. and Martinez, R.,Phys. Rev. {\bf D90}:   073001 (2014). 
			
			\bibitem{Dias:2012xp} Dias, A.G. and de S.Pires, C.A. and Rodrigues da Silva, P.S. and Sampieri, A., Phys. Rev. \textbf{D86}: 035007 (2012). 
			
			\bibitem{Hue:2015fbb} L.T. Hue, H.N. Long, T.T. Thuc, and T. Phong Nguyen, Nucl.Phys. \textbf{B907}: 37 (2016). 
			
			\bibitem{Thuc:2016qva} T.T. Thuc, L.T. Hue, H.N. Long, and T. Phong Nguyen, Phys.Rev. \textbf{D 93}: 115026 (2016). 
			
			\bibitem{Zhang:2015csm} Zhang, Hai-Bin and Feng, Tai-Fu and Zhao, Shu-Min and Yan, Yu-Li and Sun, Fei,  Chin. Phys. C, {\bf 41}: 043106 (2017). 
			
			\bibitem{Herrero-Garcia:2017xdu} Herrero-Garcia, Juan and Ohlsson, Tommy and Riad, Stella and Wiren, Jens,  JHEP {\bf 04}: 130 (2017). 
			
			\bibitem{Blankenburg:2012ex} Blankenburg, Gianluca and Ellis, John and Isidori, Gino, Phys. Lett. B {\bf 712}: 386-390 (2012). 
			
			\bibitem{Pilaftsis:1992st} Pilaftsis, Apostolos, Phys. Lett. B  \textbf{285}: 68-74 (1992). 
			
			\bibitem{Herrero-Garcia:2016uab} Herrero-Garcia, Juan and Rius, Nuria and Santamaria, Arcadi, JHEP {\bf 11}: 084 (2016). 
			
						
			\bibitem{Hong:2024yhk} Hong, T. T. and Phuong, L. T. T. and Nguyen, T. Phong and Nha, N. H. T. and Hue, L. T., Phys. Rev. D, \textbf{110} (7): 075010 (2024). 
			
			\bibitem{Hong:2024swk} Hong, T. T. and Hue, L. T. and Phuong, L. T. T. and Nha, N. H. T. and Nguyen, T. Phong, Phys. Scripta, \textbf{99}, (12):125308 (2024). 
			
			\bibitem{Duy:2024txy} Duy, N. T. and Huong, D. T. and Van Loi, Duong and Van Dong, Phung , Eur. Phys. J. C, \textbf{85}, (9): 1053 (2025). 
			
			
			\bibitem{CMS:2019pex} Sirunyan, Albert M and others, JHEP \textbf{03}: 103 (2020). 
					
			\bibitem{CMS:2021ctt} Sirunyan, Albert M and others, JHEP, \textbf{07}: 208 (2021). 
			
		 \bibitem{Dias:2003iq} Dias, Alex G. and de S. Pires, C. A. and Rodrigues da Silva, P. S., Phys. Rev. D, \textbf{68}: 115009 (2003). 
			
			\bibitem{Dias:2003zt} Dias, Alex G. and Pleitez, V., Phys. Rev. D, \textbf{69}: 077702 (2004). 
			
			\bibitem{Dias:2002gg} Dias, Alex G. and Pleitez, V. and Tonasse, M. D., Phys. Rev. D, \textbf{67}: 095008 (2003). 
				
			\bibitem{Ferreira:2016uao} Ferreira, J. G. and de S. Pires, C. A and Rodrigues, J. G. and Rodrigues da Silva, P. S., Phys. Lett. B, \textbf{771}: 199-205 (2017). 
			
						
			\bibitem{alp331} V. H. Binh, D. T. Binh,  A. E. C\'arcamo Hern\'andez, D. T. Huong, D. V. Soa, and  H. N. Long, Phys. Rev. D {\bf 107}: 095030 (2023). 
			
			\bibitem{cl} D. Chang and H. N. Long, Phys. Rev. D \textbf{73}: 053006 (2006). 			
			
			\bibitem{Long:2024rdq} Long, H. N. and Hung, H. T. and Binh, V. H. and Arbuzov, A. B., 2024,  [arXiv:2412.04269 [hep-ph]].
					
		\bibitem{Ibarra:2010xw} A. Ibarra, E. Molinaro, and S.T. Petcov, JHEP \textbf{09}: 108 (2010).
			
			
			\bibitem{il} H.  N. Long and T. Inami, Phys. Rev. {\bf D 61}: 075002 (2000).
			
		  \bibitem{Dias:2002hz} Dias, Alex G. and Pleitez, V. and Tonasse, M. D., Phys. Rev. D, \textbf{69}: 015007 (2004).
		   
		   
			\bibitem{Crivellin:2018qmi} Crivellin, Andreas and Hoferichter, Martin and Schmidt-Wellenburg, Philipp, Phys. Rev. D.\textbf{98}: 113002 (2018). 	
			
			\bibitem{Quyet:2023dlo} Quyet, H. V. and Hieu, T. T. and Tham, N. T. and Hang, N. T. T. and Hung, H. T., Chin. J. Phys.\textbf{90}: 166-186 (2024). 
				
			\bibitem{Hung:2019jue}
			Hung, H. T. and Hong, T. T. and Phuong, H. H. and Mai, H.L. T. and Hue, L. T.,
			Phys. Rev. {\bf D100}: 075014 (2019) .
			
			\bibitem{Denner:2011mq} Denner, A. and Heinemeyer, S. and Puljak, I. and Rebuzzi, D. and Spira, M., Eur. Phys. J. C, \textbf{71}: 1753 (2011). 	
			
			\bibitem{CMS:2012hfp} Chatrchyan, Serguei and others, Phys. Lett. B, \textbf{710}: 403-425 (2012). 
			
			\bibitem{CMS:2012zhx} Chatrchyan, Serguei and others, Phys. Lett. B, \textbf{710}: 26-48, (2012).	
			
			\bibitem{Barger:2012hv} Barger, Vernon and Ishida, Muneyuki and Keung, Wai-Yee, Phys. Rev. Lett., \textbf{108}: 261801 (2012). 
						
			\bibitem{JUNO:2021vlw} Abusleme, Angel and others, Prog. Part. Nucl. Phys.\textbf{123}: 103927 (2022). 
			
			
			\bibitem{Hue:2021zyw} Hue, L. T. and Phan, Khiem Hong and Nguyen, T. Phong and Long, H. N. and Hung, H. T., Eur. Phys. J. C.\textbf{82}, (8):722 (2022). 
			
						
			\bibitem{LHCHiggsCrossSectionWorkingGroup:2013rie} Andersen, J R and others, CERN-2013-004, month 7, 2013, [arXiv:1307.1347[hep-ph]].
			
			\bibitem{Langenfeld:2012ti} Langenfeld, Ulrich and Moch, Sven-Olaf and Pfoh, Torsten, JHEP \textbf{11}: 070 (2012). 
						
			\bibitem{Das:2025tuk} Das, Goutam and Moch, Sven-Olaf, TTK-25-34, P3H-25-080, doi: 10.21468/SciPostPhysCommRep.17, 2025, [arXiv:2510.17413 [hep-ph]].
			
		    \bibitem{Anastasiou:2016cez} Anastasiou, Charalampos and Duhr, Claude and Dulat, Falko and Furlan, Elisabetta and Gehrmann, Thomas and Herzog, Franz and Lazopoulos, Achilleas and Mistlberger, Bernhard, JHEP \textbf{05}: 058 (2016). 
		    
		    \bibitem{Davies:2019wmk} Davies, Joshua and Herren, Florian and Steinhauser, Matthias, Phys. Rev. Lett. \textbf{124} (11): 112002 (2020).  
	
	      \bibitem{Harlander:2010su} Robert V. Harlander et al, Eur.Phys.J C \textbf{66}: 359-372 (2010). 
	
			\bibitem{Ninh:2005su} Ninh, Le Duc and Long, Hoang Ngoc, Phys. Rev. D, \textbf{72}: 075004 (2005). 			
						
			\bibitem{Perez:2004jc} Perez, M. A. and Tavares-Velasco, G. and Toscano, J. J., Phys. Rev. D, \textbf{69}, p 115004, 2004, [arXiv:0402156 [hep-ph]].
			
				
			\bibitem{Athron:2015tsa} Athron, P. and Harries, D. and Williams, A. G., Phys. Rev. D, \textbf{91}: 115024 (2015). 
			
			\bibitem{Cvetic:1997mbh} Cvetic, Mirjam and Langacker, P., Adv. Ser. Direct. High Energy Phys., \textbf{18}: 312--331 (1998). 
			
			\bibitem{Cannoni:2015wba} Cannoni, Mirco, Eur. Phys. J. C, \textbf{76} (3): 137 (2016). 
				
			\bibitem{Bertone:2004pz} Bertone, Gianfranco and Hooper, Dan and Silk, Joseph, Phys. Rept., \textbf{405}: 279-390 (2005). 
			
			\bibitem{Planck:2018vyg} Aghanim, N. and others, Astron. Astrophys., \textbf{641}: A6 (2020). 	
			
			\bibitem{Bahcall:1999xn} Bahcall, Neta A. and Ostriker, Jeremiah P. and Perlmutter, Saul and Steinhardt, Paul J., Science, \textbf{248}: 1481-1488 (1999). 
					
			\bibitem{Passarino:1978jh} Passarino, G. and Veltman, M. J. G., Nucl. Phys. B \textbf{160}: 151-207 (1979). 
			
			
			\bibitem{Phan:2016ouz} Phan, K.H. and Hung, H.T. and Hue, L.T., PTEP.\textbf{2016} (11): 113B03 (2016). 
						
			\bibitem{Hue:2017lak} Hue, L. T. and Ninh, L. D. and Thuc, T. T. and Dat, N. T. T., Eur. Phys. J. C, vol.\textbf{78} (2): 128 (2018). 
			
			\bibitem{Hue:2010xr} Hue, L. T. and Hung, H. T. and Long, H. N., Rept. Math. Phys.\textbf{69}: 331-351 (2012). 
						
			\bibitem{Dong:2009db} Dong, P. V. and Hue, L. T. and Hung, H. T. and Long, H. N. and Thao, N. H., Theor. Math. Phys.\textbf{165}: 1500-1511 (2010).
				
		\end{thebibliography}
	\end{document}